\documentclass[preprint,aps]{revtex4}
\usepackage[latin9]{luainputenc}
\usepackage{float}
\usepackage{amsmath}
\usepackage{amssymb}
\usepackage{graphicx}
\usepackage{xcolor}
\usepackage{setspace}
\usepackage[unicode=true,pdfusetitle,
bookmarks=true,bookmarksnumbered=false,bookmarksopen=false,
breaklinks=false,pdfborder={0 0 1},backref=false,colorlinks=false]
{hyperref}
\DeclareMathOperator{\arctanh}{arctanh}
\makeatletter

\@ifundefined{textcolor}{}
{%
	\definecolor{BLACK}{gray}{0}
	\definecolor{WHITE}{gray}{1}
	\definecolor{RED}{rgb}{1,0,0}
	\definecolor{GREEN}{rgb}{0,1,0}
	\definecolor{BLUE}{rgb}{0,0,1}
	\definecolor{CYAN}{cmyk}{1,0,0,0}
	\definecolor{MAGENTA}{cmyk}{0,1,0,0}
	\definecolor{YELLOW}{cmyk}{0,0,1,0}
}

\usepackage{amsfonts}
\usepackage{bm}
\usepackage[dvips]{epsfig}
\usepackage[dvips]{graphicx}
\usepackage{float}
\usepackage{dcolumn}
\usepackage{ifpdf}
\usepackage{dcolumn}
\usepackage{multirow}

\newcommand{\be}{\begin{equation}}
\newcommand{\ee}{\end{equation}}
\newcommand{\bes}{\begin{subequations}}
	\newcommand{\ees}{\end{subequations}}
\newcommand{\ben}{\begin{eqnarray}}
\newcommand{\een}{\end{eqnarray}}

\makeatother

\begin{document}
\title{Semi-compactness and multiple oscillating pulses in kink scattering}
 \author{D. Bazeia$^1$, Adalto R. Gomes$^{2}$, Fabiano C. Simas$^{2,3}$}
 \email{bazeia@fisica.ufpb.br, argomes.ufma@gmail.com, adalto.gomes@ufma.br, fc.simas@ufma.br}
  \noaffiliation
\affiliation{
$^1$ Departamento de F\'isica, Universidade Federal da Para\'iba, 58051-970, Jo\~ao Pessoa, PB, Brazil\\
$^2$ Programa de P\'os-Gradua\c c\~ao em F\'\i sica, Universidade Federal do Maranh\~ao (UFMA),
Campus Universit\'ario do Bacanga, 65085-580, S\~ao Lu\'\i s, Maranh\~ao, Brazil\\
$^3$ Centro de Ci\^encias Agr\'arias e Ambientais-CCAA, Universidade Federal do Maranh\~ao
(UFMA), 65500-000, Chapadinha, Maranh\~ao, Brazil
}
\noaffiliation

\begin{abstract}
In this work we consider model of asymmetric kinks, where the behavior of the solution in one side is different from the other side. Also, the models depend of an integer $n$ and, with the increase of $n$, the constructed kink assumes a hybrid character: a compactlike profile on one side and a kinklike profile on the other side. We investigate numerically the kink-antikink and antikink-kink dynamics, with the aim to understand the effect of the transition of the usual kink to the semi-compacton structure.  The kink-antikink process shows the formation of one-bounce windows for small values of $n$. The increase of $n$ favors the breaking this structure and the appearance of oscillatory modes. For antikink-kink collisions we report the appearance of two-bounce windows for small values of the parameter. We also found an intricate structure of two-oscillation windows.

\end{abstract}

\pacs{ XXXXXX }


\maketitle


\section{ Introduction }


Solitons are solutions of nonlinear theories with special properties, such as topological profile, concentrated energy density, and finite energy; also, they maintain their form even after scattering with another soliton solution. More generally, solitary waves have concentrated energy density and finite energy, but they may not have topological profile and may interact changing form. They appear in several contexts in nonlinear physics, including optics \cite{abdu}, condensed matter \cite{dau, bisho} and high energy physics \cite{manton, vascha}. As examples of solitary waves that may or may not be solitons, we have kinks, lumps, vortices, magnetic monopoles and skyrmions \cite{manton}. Kinks are perhaps the simplest of these solutions, and they in general are not solitons, although kinks in the sine-Gordon model are good and simple examples of solitons. In fundamental physics, localized solutions are nicely considered in braneworld cosmology \cite{bucher,lang,trodd,maeda}. Effects of small nonplanar fluctuations in the collision of planar domain walls in $(3,1)$ dimensions were investigated in the Ref. \cite{bbh}. Planar approximation for bubble collision is justifiable for nucleated bubbles that has been expanded considerably before colliding \cite{bbh}. Bubble collisions in first-order phase transitions in the ultrarelativistic limit were considered in the Ref.  \cite{jinno}. Furthermore, domain walls have been used as a possibility to generate the Big Bang conditions \cite{kho,maeda1}.

Kinks (as well as the corresponding antikinks) in (1,1) dimensions are very interesting localized structures, and have been studied in several distinct cases, in particular the scattering of kinks and antikinks have been intensively investigated in the recent literature. In integrable models like sine-Gordon \cite{cue} and KdV \cite{fermi} the kink-antikink pair collide elastically and acquire at most a phase shift.  However, kink scattering in non-integrable system is rich and have several surprising effects. In particular, a well studied nonintegrable model is the $\phi^4$ model \cite{sugi,moshi,win1,csw,belo1,aninos,good1}. The intricate kink-antikink ($K \bar K$) scattering process depends crucially on the initial velocity. In this sense, a critical value of velocity ($v_c$) separates two regions: i)  inelastic scattering for $v>v_c$, ii) formation of a bion state, for $v<v_c$. There, a composited kink-antikink state radiates until the complete annihilation of the pair. An intriguing behavior is observed for some intervals of $v\lesssim v_c$. In these intermediate values of velocities there is the formation of two-bounce windows, where the kink-antikink pair collide twice before receding each other. The appearance of such  resonance windows is related to the energy exchange mechanism between the translational and vibrational modes \cite{csw}.

Kink scattering has been investigated in several models. One can cite the polynomials with one scalar field, for instance, modified $\phi^4$ model \cite{adalto4,adalto5,moray}, $\phi^6$ model \cite{dorey1,gani1,adalto6}, $\phi^8$ model \cite{gani2,manton2,gani3}, nonpolynomial \cite{campbell,gani4,gani5,bazeia4}, models which possess kinks with power-law tails \cite{chri,chri1,chri2} and models with two scalar fields \cite{hala,alonso1,alonso2,alonso3,alonso4,alonso5}. Furthermore, there are studies of kink collision with a boundary \cite{dorey2,dorey3,adalto1}, with impurities \cite{good2,fei,roy,joao,adam2}, multi-kink collision \cite{marja,marja1,marja2,gani6} and other models \cite{adalto2, adalto3,mendoca,mendoca1,adalto7}.

In general, the presence of internal (vibrational)  mode is an essential ingredient to understand the presence of two-bounce windows. However, in the Refs. \cite{dorey1,gani3}, two-bounce windows were found despite the absence of an internal oscillatory mode. This can be understood if we consider the perturbation of the antikink-kink pair \cite{dorey1,gani3,gani7}. In \cite{adalto4}, the deformed $\phi^4$ model leads to appearance of more than one vibrational mode. There one can observe the total suppression of two-bounce windows even with the presence of internal mode. In the Refs. \cite{dorey4,joao2}, it was investigated the role of quasinormal modes in kink scattering. The Ref. \cite{dorey4} shows that the quasinormal modes can store energy during the collision. However, these modes decay exponentially which leads to energy leakage. A similar behavior is observed in Ref. \cite{joao2}, where the square well potential is modified allowing the appearance of quasinormal modes. The effect of internal modes was also studied in the Refs. \cite{adam1,adam2}. The Ref. \cite{adam2} shows the importance of oscillatory modes in the formation of spectral walls. This phenomenon occurs after that the internal mode disappears into continuous spectrum. Moreover, in the Ref. \cite{liu} is analyzed the kink-antikink collision in a  Lorentz-violating $\phi^4$ model. In that work, the influence of the Lorentz term change the localization of the two-bounce windows. Furthermore, the radiation emitted by the kink-antikink is asymmetric, related by the birefringence phenomenon.

Compactons were proposed and studied in the Ref. \cite{phil} considering a  Korteweg-de Vries-like equation with nonlinear dispersion.  The compacton has energy density that vanishes outside of compact space. On the other hand, the energy density of the kink vanishes asymptotically. The Ref. \cite{sacco} analyzed the relevance of nonlinear stacking potential in models of double-stranded DNA. They showed that this interaction has a fundamental role in the location of energy along the biomolecule. Moreover, the choice of this potential provides the appearance of compactons. Compact baby skyrmions with nontrivial topological charge were obtained in the Ref. \cite{adam}. The Ref. \cite{bazeia6} presented a route to transform kink solutions in compacton structures. Moreover, the authors observed the behavior of this process in a braneworld scenario, revealing a symmetric hybrid brane configuration. In the Ref. \cite{bazeia1} an asymmetric compact structure was built in which the scalar field generates an asymmetric hybrid brane. Recently, have been discussed the scattering of generalized $\phi^4$ model that support kinks with a compact profile \cite{bazeia8}. In particular, the authors show the emergence of metastable structures with the formation of compact defects. Moreover, for small velocities, are formed long-lived oscillating structure, named compactlike oscillons. The scattering of generalized compact oscillons in the signum-Gordon model were studied in the Ref. \cite{hahne}, showing the generation of oscillons in the form of jet-like cascates.

Oscillons appear in primordial cosmology models, mostly related with bubble collisions \cite{oscill1} and the inflationary paradigm \cite{oscill2}, but also with string cosmology \cite{oscill3}. It is of special interest to identify the conditions that allow the production of oscillons \cite{oscill4, oscill5}. Some works analyze the time evolution and decay of these solutions \cite{oscill6, oscill7}. Then it is of interest to consider other aspects of models that lead the production of oscillons, as noted in the kink-antikink scattering \cite{gani5,bazeia4} and multi-kink collisions \cite{gani6}.

In this work we consider the collision of defects that have a semi-compacton character. As far as we know, scattering of this type of defects has not been reported before. In the section II we consider the asymmetric compactlike structures proposed in the Ref. \cite{bazeia1}. We analyze the spectrum of linear perturbations for isolated kink/antikink solutions and the collective kink-antikink and antikink-kink pair. In section III we present our results for kink-antikink and antikink-kink collisions. In both cases we report the appearance of oscillating pulses forming a structure of windows, whose richness can be related to the semi-compactness of the solutions. Finally, we conclude the present work in section IV.


\section { The Model }

\begin{figure}
	\includegraphics[{angle=0,width=8cm,height=5cm}]{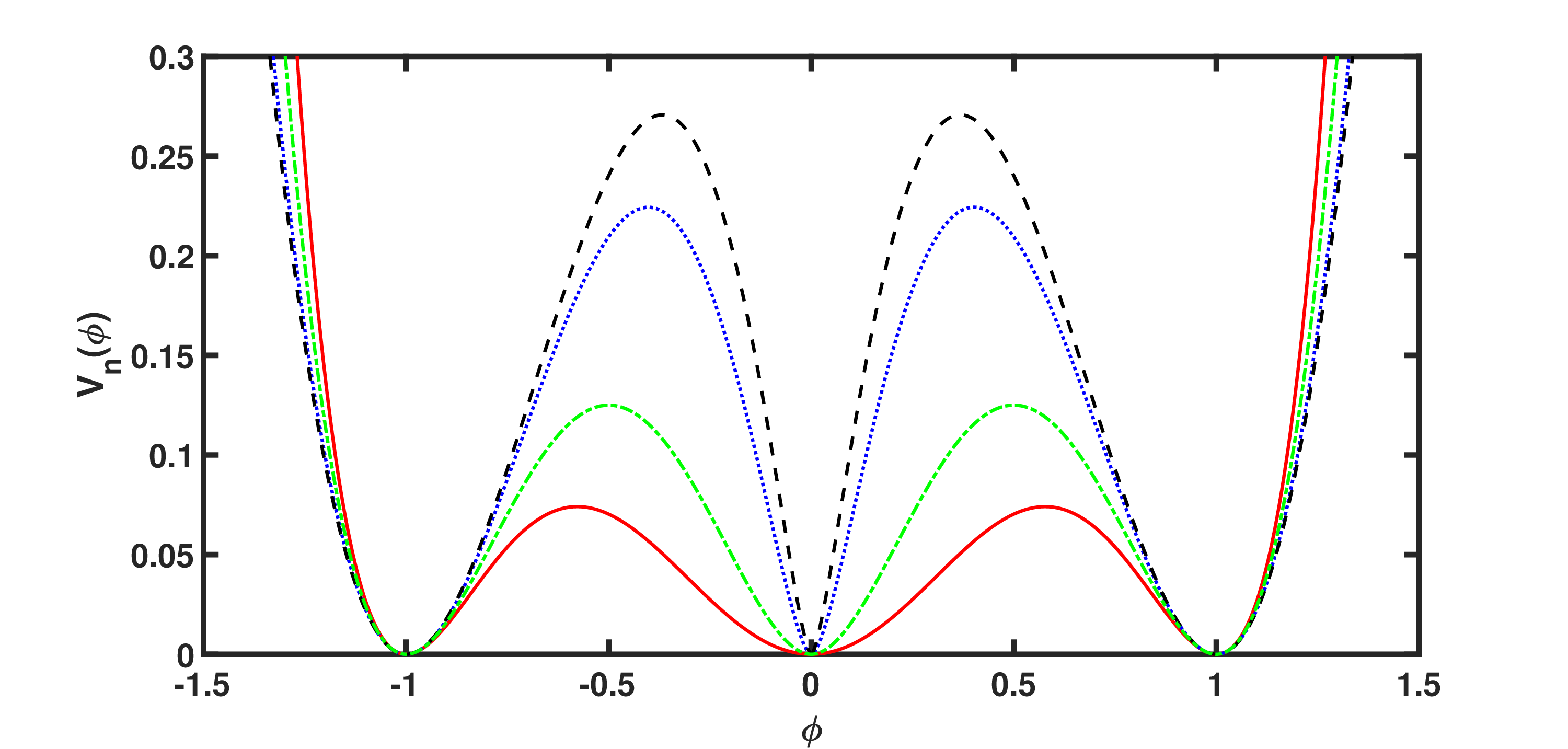}
	\includegraphics[{angle=0,width=8cm,height=5cm}]{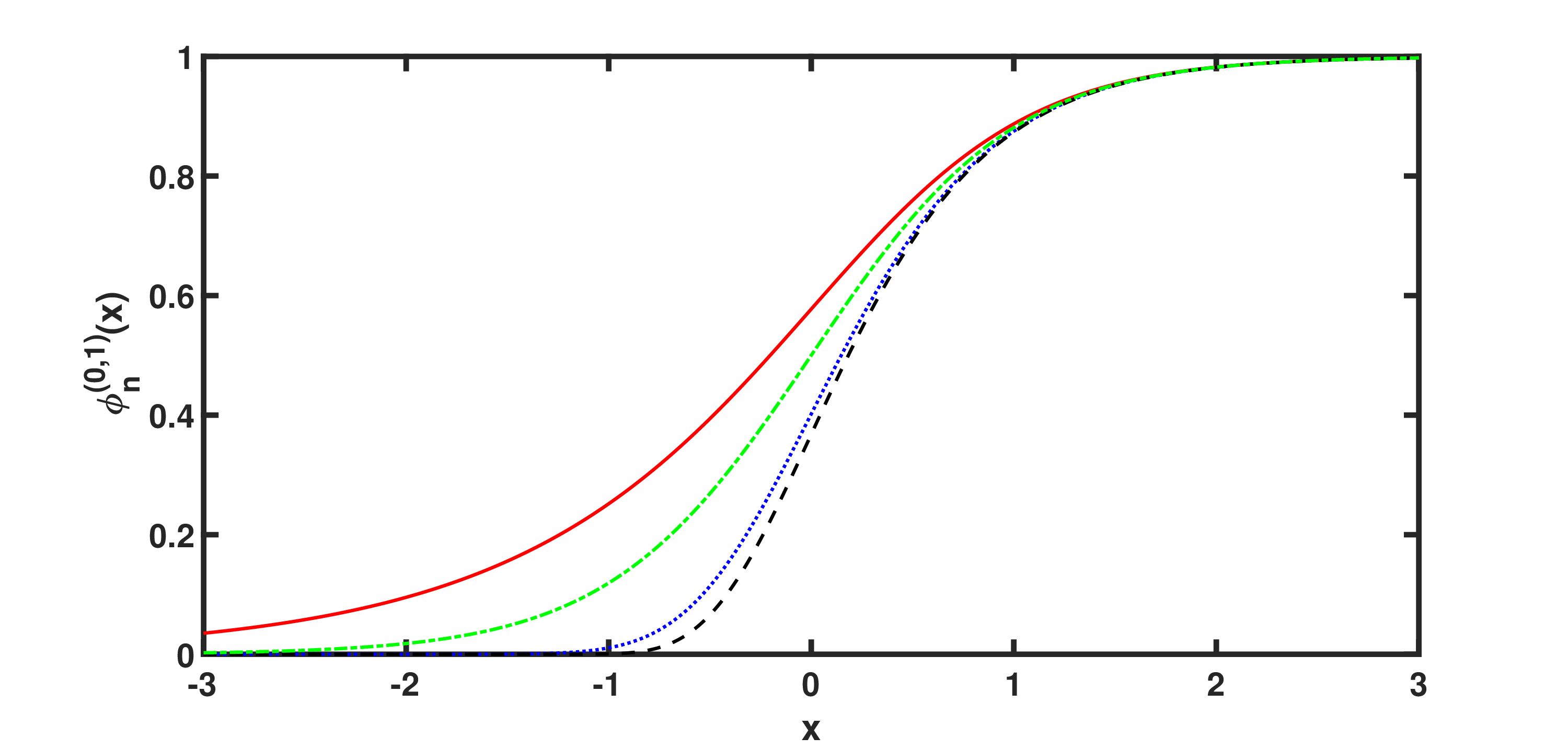}
	\caption{(a) Potential $V(\phi)$ and (b) field configurations $\phi_n^{(0,1)}(x)$ for $n=1$ (red solid), $n=2$ (green dash-dotted)  $n=10$ (blue dotted) and $n \rightarrow \infty $ (black dash).}
		\label{fig1}
	\end{figure}

We consider the following action with standard dynamics
\be
S=\int dt dx \bigg( {\frac12 \partial_{\mu} \partial^{\mu} \phi - V(\phi)} \bigg),
\label{action}
\ee
where  the potential is given by \cite{bazeia1}

\be
V_n(\phi) = \frac12 n^2 \phi^2 [1-(\phi^2)^{\frac1n}]^2,
\label{pot}
\ee
and $n$ is a positive integer.  For $n=1$ we recover the $\phi^6$ model,
\be
V_1=\frac12 \phi^2 (1-\phi^{2})^2.
\ee
whereas for $n\to\infty$ we have
\be
V_{\infty}=\frac12\phi^2\ln^2(\phi^2).
\ee
The Fig. \ref{fig1}a shows the plots of $V_n(\phi)$ for some values of $n$. Note that the potential is $Z_2$ symmetric and has three minima $\{-1,0,1\}$ and two local maxima that 
 increase with  $n$. Then we have two topological sectors, and without loosing generality we will consider only the sector $0 \leq \phi \leq 1$.

The equation of motion is given by
\be
\frac{\partial^2 \phi}{\partial t^2}-\frac{\partial^2 \phi}{\partial x^2} + n^2\phi[1-(\phi^2)^\frac1n]^2 - 2n\phi[1-(\phi^2)^\frac1n] (\phi^2)^\frac1n=0.
\label{eqm}
\ee
and the static kink solution in the sector $0 \leq \phi \leq 1$ is given by \cite{bazeia1}
\be
\phi_n^{(0,1)} (x) = \Bigg[  \frac{1+\tanh\Big(x-a \Big) }{2} \Bigg]^{\frac n2},
\label{exp}
\ee
where 
\be 
a= \arctanh\big(\frac{2-n}{2+n}\big)
\ee
is a specific constant of integration chosen to guarantee that the local maximum of the potential $V(\phi)$ occurs at $\phi= \phi_n^{(0,1)} (0)$ \cite{bazeia1}. The antikink solutions are given by $\phi_n^{(1,0)}(x) = \phi_n^{(0,1)}(-x)$. The solution connecting the minima $\{-1,0\}$ can be found by taking $\phi_n^{(-1,0)}(x) = - \phi_n^{(0,1)}(x)$. The Fig. \ref{fig1}b shows the plots of the kink profile $\phi_n{(0,1)}$ for the same values of $n$ considered in the Fig. \ref{fig1}a. Note that the increase of $n$ gives solutions that approach faster the minimum $\phi=0$. Indeed we have, for $x\lesssim -1.5$ and $1\leq n \leq 6$,

\be
\phi_n^{(0,1)}(x) \simeq \bigg(\frac{n}{2} \bigg)^{n/2} e^{nx}.
\ee
For larger values of $n$, this approximation is good only for even lower values of $x$. On the other hand, the way the solution approach the minimum $\phi=1$ is independent of $n$. There we have, for $x\gtrsim 1.5$,

\be
\phi_n^{(0,1)}(x) \simeq 1-e^{-2x}.
\ee
An interesting limit is given for $n \to \infty$, where one has, for all values of $x$ \cite{bazeia1},

\be
\lim_{n\to\infty} \phi_n^{(0,1)}(x) = e^{-e^{-2x}}
\ee
That is, the increase of $n$ tends to form a half-compacton. Note that for $n\to\infty$, the scalar field is already too close from the minimum $\phi=0$ at a finite value of $x<0$. In this case the solution is compact-like for $x<0$ and kink-like for $x>0$. Note also from the Fig. \ref{fig1}b that the behavior of the scalar field in the region $x>0$ is almost independent of $n$ for large $x$. That is, the main influence of $n$ is in the region $x<0$ and in the vicinity of $x=0$ (up to $x=1.5$ in the scale of Fig. \ref{fig1}b). Close to the vacuum $\phi=1$, the solutions are roughly the same, independent of $n$.

\begin{figure}
	\includegraphics[{angle=0,width=8cm,height=5cm}]{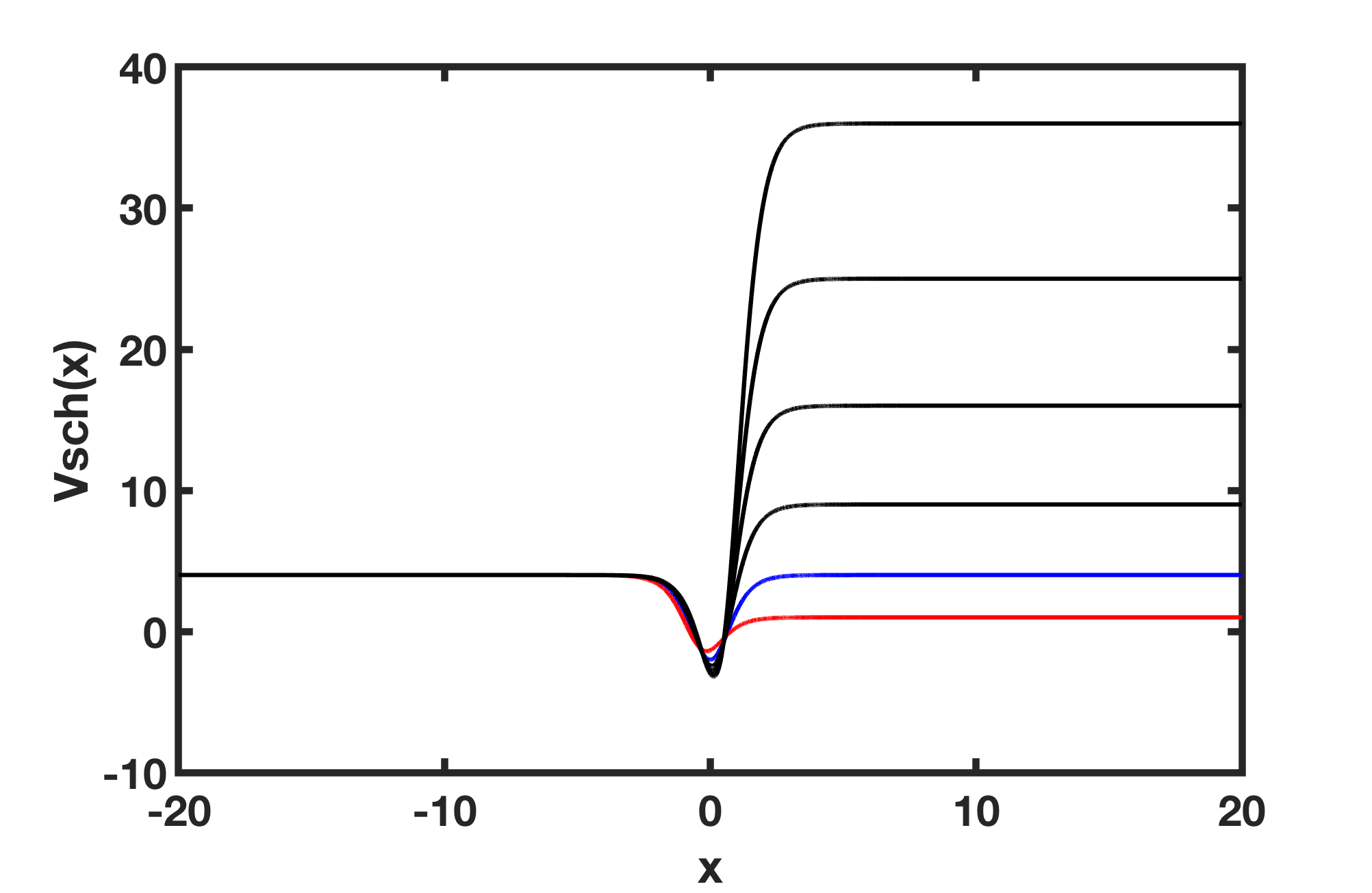}
	\includegraphics[{angle=0,width=8cm,height=5cm}]{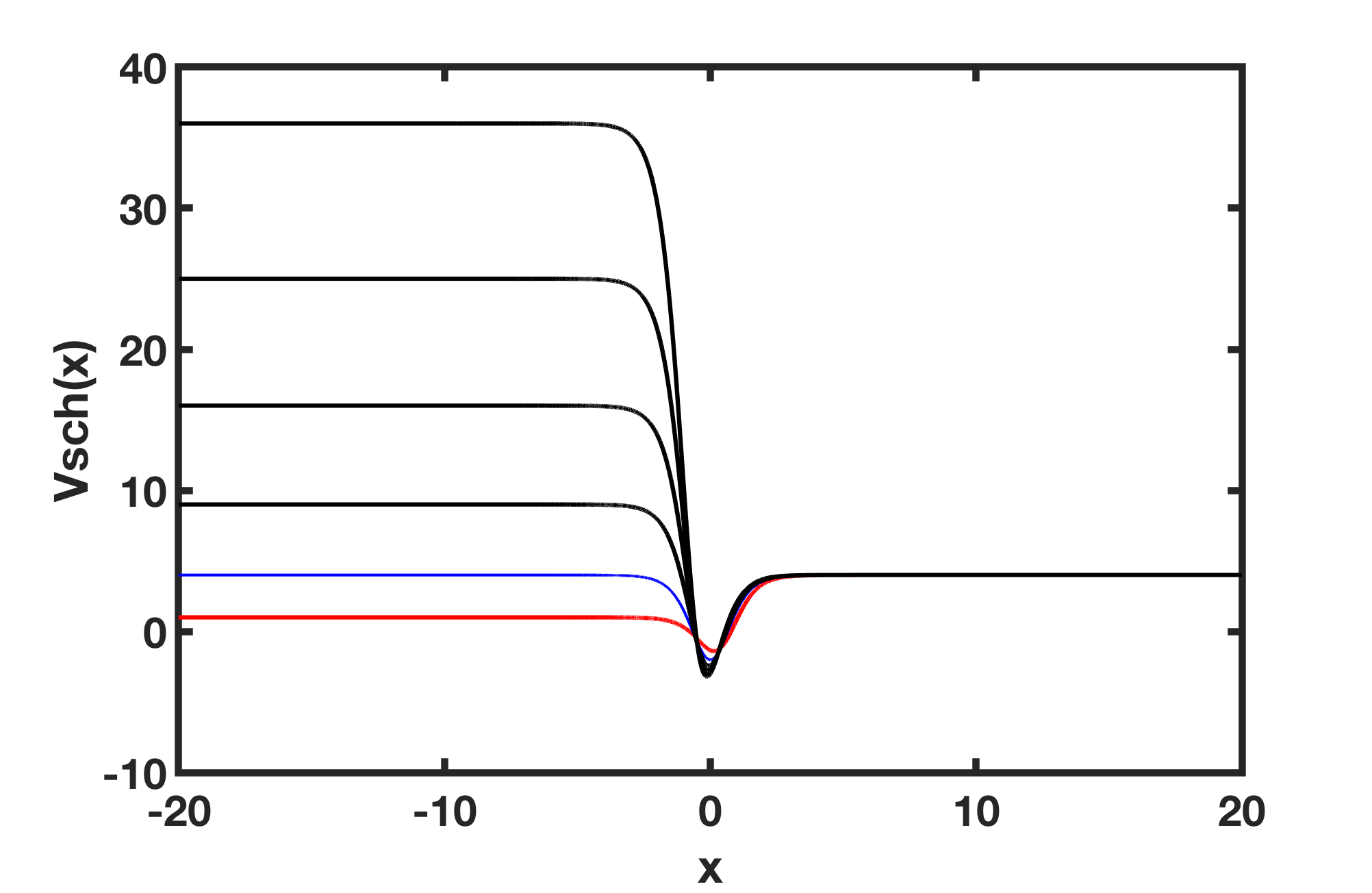}
	\caption{The Schr\"odinger potential $V_{sch}(x)$ for (a) kink $\phi_n^{(0,1)}$ and (b) antikink $\phi_n^{(1,0)}$. In all figures we fixed $n=1$ (red solid), $n=2$ (blue solid) and $3 \leq n \leq 6$ (black line).}
	\label{fig2}
\end{figure}

Perturbing linearly the scalar field around the static kink/antikink solution $\phi_s(x)$ as $\phi(x,t) = \phi_s(x) + \eta(x) \cos(\omega t)$, we arrive at a Schr\"odinger-like equation 
\be
-\frac{d^2 \eta(x)}{dx^2} + V_{sch}(x) \eta(x) = \omega^2 \eta(x),
\label{sch}
\ee
with  $V_{sch}(x) = V_{\phi \phi}(\phi_s(x))$.  The Schr\"odinger-like potentials for kink ($\phi_n^{(0,1)}$) and antikink ($\phi_n^{(1,0)}$) solutions are depicted in the Fig. \ref{fig2}a and Fig. \ref{fig2}b, respectively. We see that $V_{sch}(x\to -\infty)$ (for the kink) and $V_{sch}(x\to +\infty)$ (for the antikink), increases with $n$. Indeed, from Eq. (\ref{pot}) we have for the squared mass $m_n^2$ of the elementary excitations at the vacuum  $\phi=0$,   $m_n^2=V_{\phi\phi}|_{\phi=0}=n^2$. The asymptotic values of the kink stability potential correspond to $V_{sch}(x)=n^2$ for $x\rightarrow -\infty$ and $V_{sch}(x)=4$ for $x\rightarrow + \infty$. Note also that $V_{sch}(x)$ for the kink corresponds to $V_{sch}(-x)$ for the antikink. 

\begin{figure}
	\includegraphics[{angle=0,width=10cm,height=5cm}]{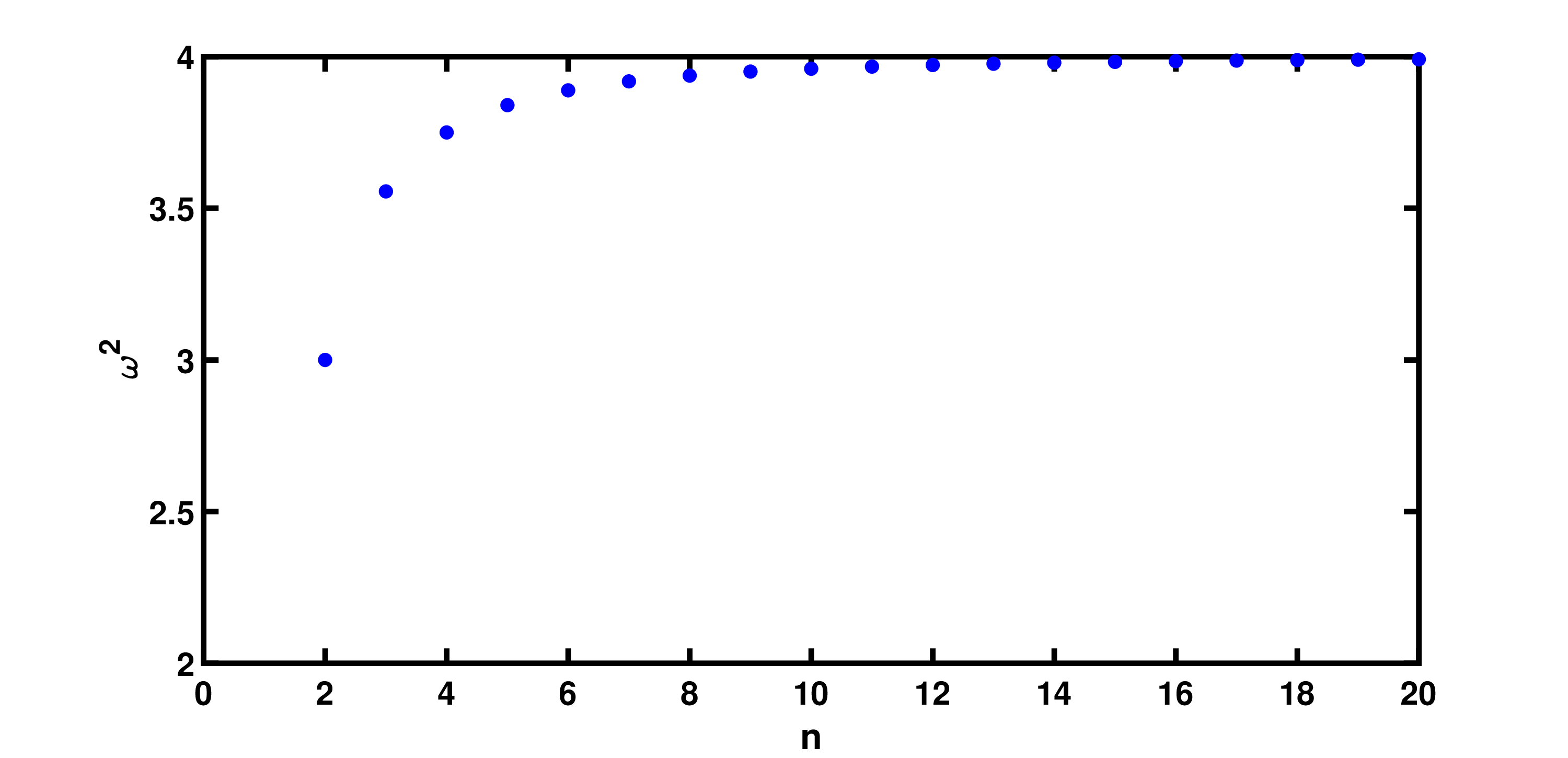}
	\caption{The squared frequencies $\omega^2$ of the vibrational states as a function of parameter $n$ for an isolated kink or antikink.}
	\label{fig3}
\end{figure}

We solved numerically the Eq.  (\ref{sch}) for a single kink or antikink. For $n\geq 1$ there is always a zero-mode, responsible for the translational invariance of the solutions. For $n=1$ there is no vibrational mode, as already known for the $\phi^6$ model \cite{dorey1}. For $n\geq 2$ there is always one vibrational mode. The Fig. \ref{fig3} shows that the squared frequency of the vibrational mode increases monotonically with $n$.  

\begin{figure}
	\includegraphics[{angle=0,width=8cm,height=5cm}]{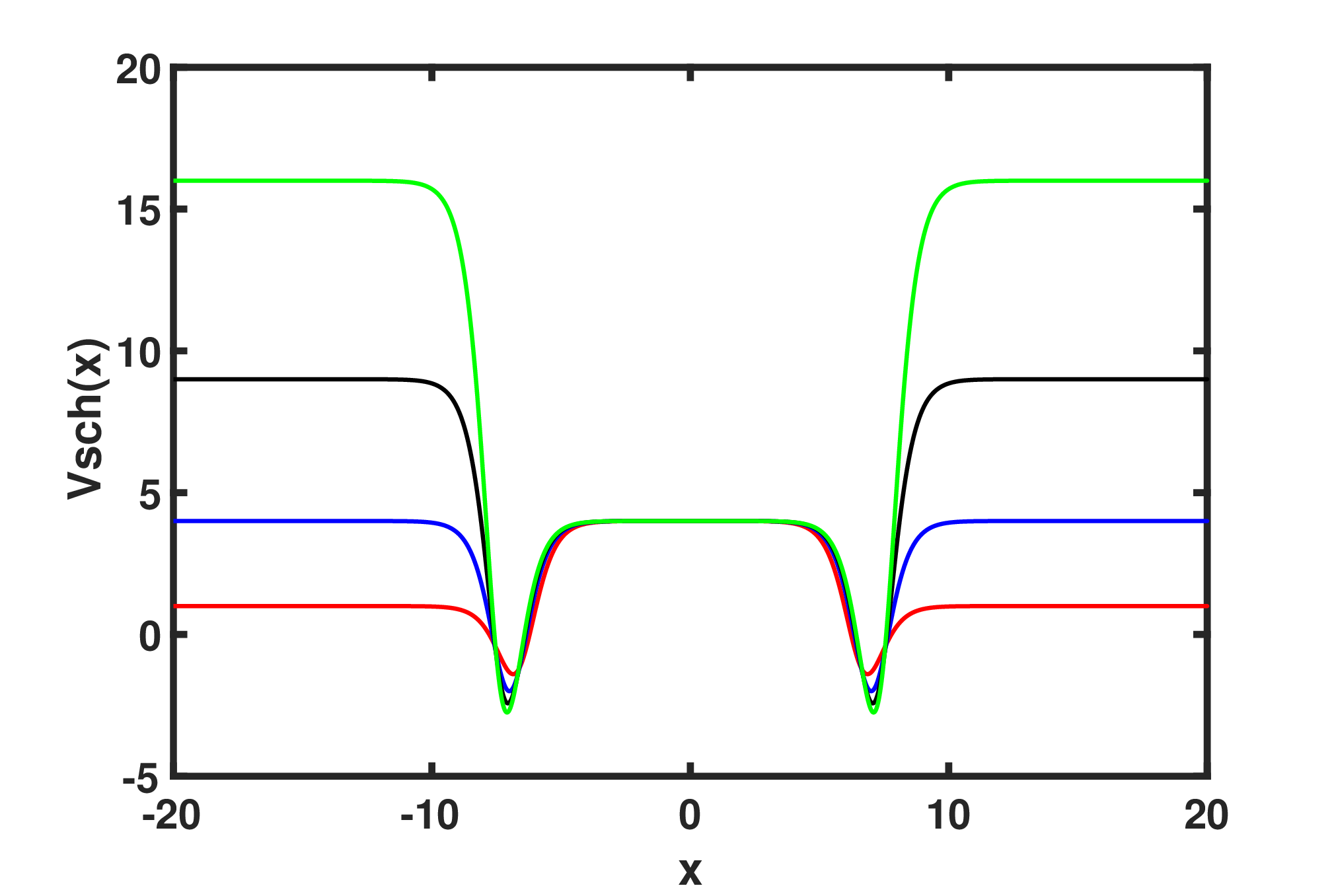}
    \includegraphics[{angle=0,width=8cm,height=5cm}]{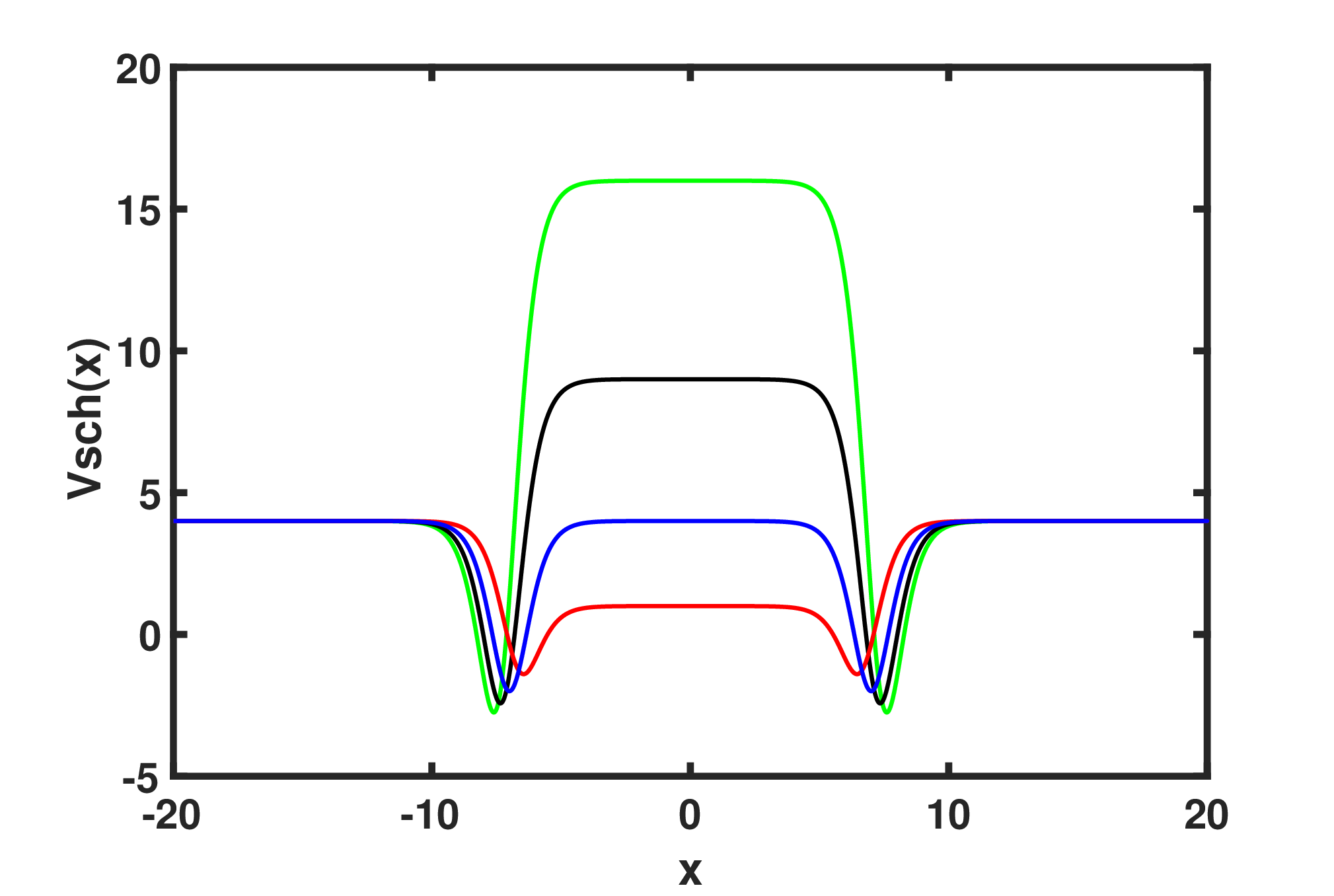}
	\caption{Schr\"odinger potential $V_{sch}(x)$ for (a) kink-antikink $\phi_n^{(0,1)}(x+x_0)+\phi_n^{(1,0)}(x-x_0)-1$ and (b) antikink-kink $\phi_n^{(1,0)}(x+x_0)+\phi_n^{(0,1)}(x-x_0)$. In all figures $x_0=8$. We fixed $n=1$ (red solid), $n=2$ (blue solid), $n=3$ (black line) and $n=4$ (green line).}
	\label{schcol}
\end{figure}

\begin{figure}
	\includegraphics[{angle=0,width=8cm,height=5cm}]{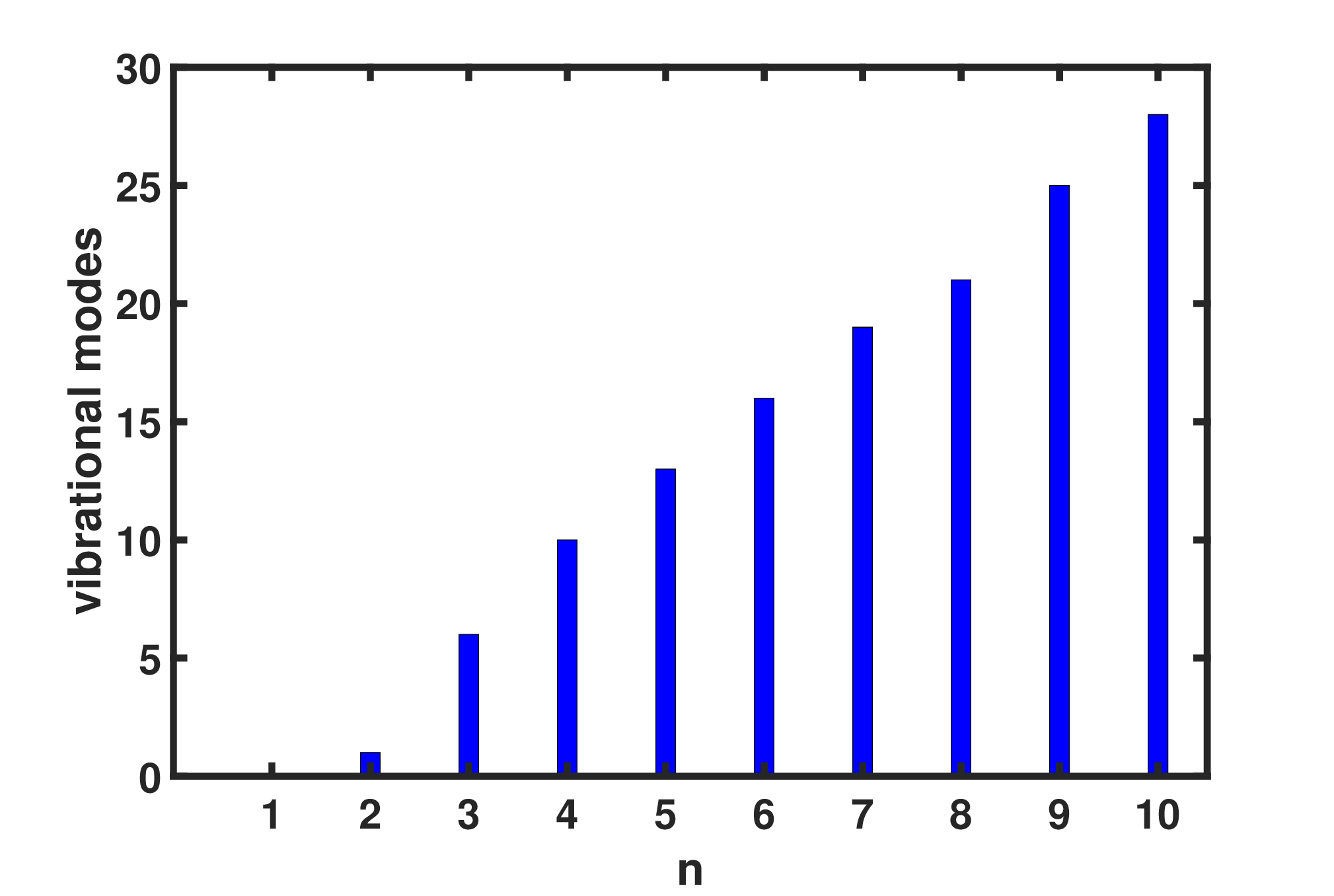}
	\includegraphics[{angle=0,width=8cm,height=5cm}]{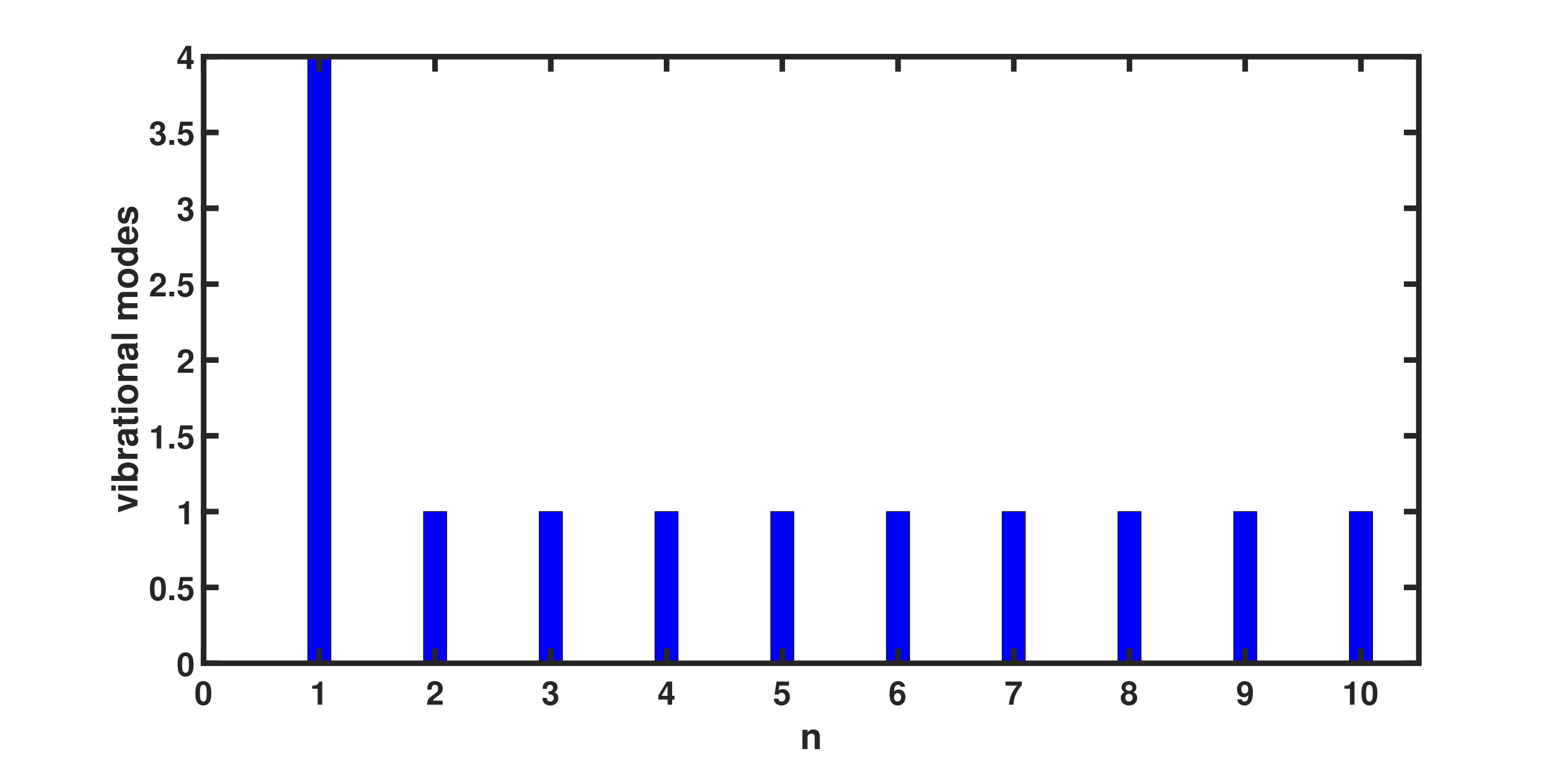}
	\caption{Number of vibrational states for (a) kink-antikink and (b) antikink-kink, corresponding to the Schr\"odinger potentials of Figs \ref{schcol}a-b.}
	\label{vibra}
\end{figure}

We also analyzed numerically the occurrence of vibrational states considering kink-antikink and antikink-kink pairs. 

The Fig. \ref{schcol}a shows the  Schr\"odinger-like potential for the kink-antikink pair for some values of $n$. The pair is separated by the distance $2x_0$, with $x_0=8$. Note from the figure that the potential is characterized by two minima at the positions of the centers of the kink and antikink. The value of the minima decrease slightly with $n$. In between the minima, the potential acquires a plateau independent of $n$. The asymptotic values at $x\to\pm\infty$ increases with $n$.  The number of vibrational modes for each value of $n$ is depicted in the  Fig. \ref{vibra}a. Note that for $n=1$ there is no vibrational mode for the kink-antikink pair, as known for the $\phi^6$ model. It has only one mode for $n=2$ and increases significatively with $n$ for $n\geq 3$. The growing of $x_0$ leads to the increasing in the number of vibrational states for $n\geq 3$, whereas the cases $n=1$ and $n=2$ is independent of $x_0$, for the kink-antikink pair. The choice of $x_0$ must guarantee that there is not a sensible overlap between the kink and antikink solutions. This is a necessary condition for the linear stability analysis.

The Fig. \ref{schcol}b shows the  Schr\"odinger-like potential for the antikink-kink pair for some values of $n$. We also considered $x_0=8$. The potential has two minima at the positions of the antikink and kink, similarly to the observed for the kink-antikink pair. The central plateau between the minima increases with $n$, whereas the asymptotic value at $x\to\pm\infty$ is constant, independent of $n$. The number of vibrational modes for each value of $n$ is depicted in the  Fig. \ref{vibra}b. Note from the figure that for $n=1$ there are four vibrational modes. These modes were considered in the Ref. \cite{dorey1} for explaining the occurrence of two-bounce windows in the $\phi^6$ model. For each $n\geq 2$ there is only one vibrational mode. The growing of $x_0$ leads to the increasing in the number of vibrational states for $n=1$,  whereas the cases  $n\geq 2$ is independent of $x_0$, since we are now dealing with antikink-kink pair.

Note that, despite the Schr\"odinger-like potential for the kink and antikink are related by a $x\to -x$ transformation, there is no such symmetry for kink-antikink and antikink-kink pairs. This means that the scattering process for the pairs must be considered separately, as we will do in the following section.


\section { Numerical Results }


Here we will discuss our main results of the kink-antikink and antikink-kink scattering in the sector $\{0,1\}$. We solved the equation of motion with a $4^{th}$ order finite-difference method with a spatial step $\delta x = 0.05$. We fixed $x_0=12$ for the initial symmetric position of the pair. For the time dependence we used a $6^{th}$ order sympletic integrator method, with a time step $\delta t=0.02$.


\subsection { Kink-antikink scattering}


For solving the equation for kink-antikink  scattering we used the following initial conditions

\begin{eqnarray}
	\phi(x,0) & = & \phi_n^{(0,1)}(x+x_0,0) + \phi_n^{(1,0)}(x-x_0,0) - 1 \\
	\dot{\phi}(x,0) & = & \dot{\phi}_n^{(0,1)}(x+x_0,0) + \dot{\phi}_n^{(1,0)}(x-x_0,0),
\end{eqnarray}
 where $\phi_n^{(0,1)}(x,t) \equiv \phi_n^{(0,1)}(\gamma(x-vt))$ means a boost of Lorentz for the static kink solution $\phi_n^{(0,1)}(x)$ and $\gamma=(1-v^2)^{-1/2}$. 
\begin{figure}
	\includegraphics[{angle=0,width=8cm,height=5cm}]{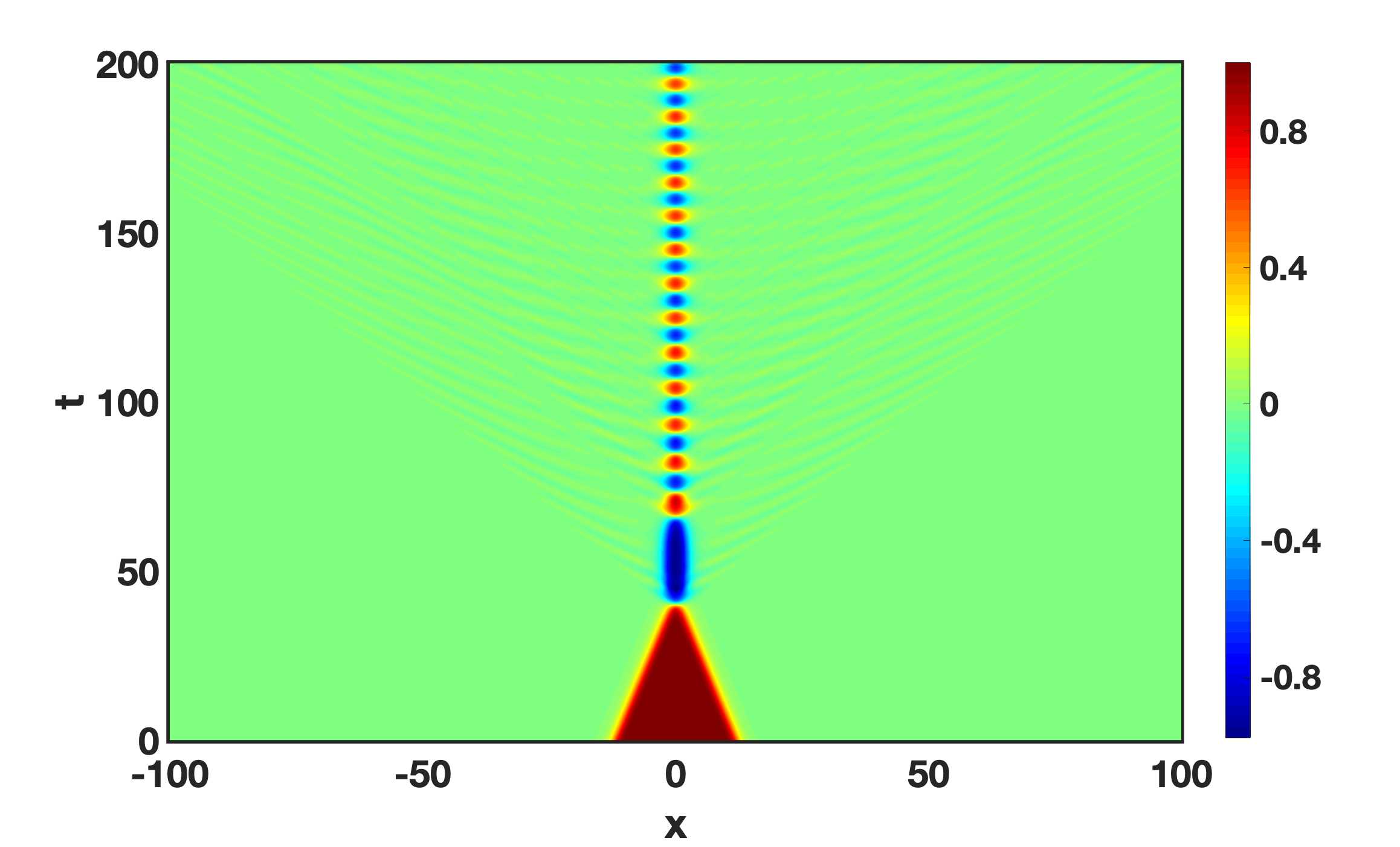}
	\includegraphics[{angle=0,width=8cm,height=5cm}]{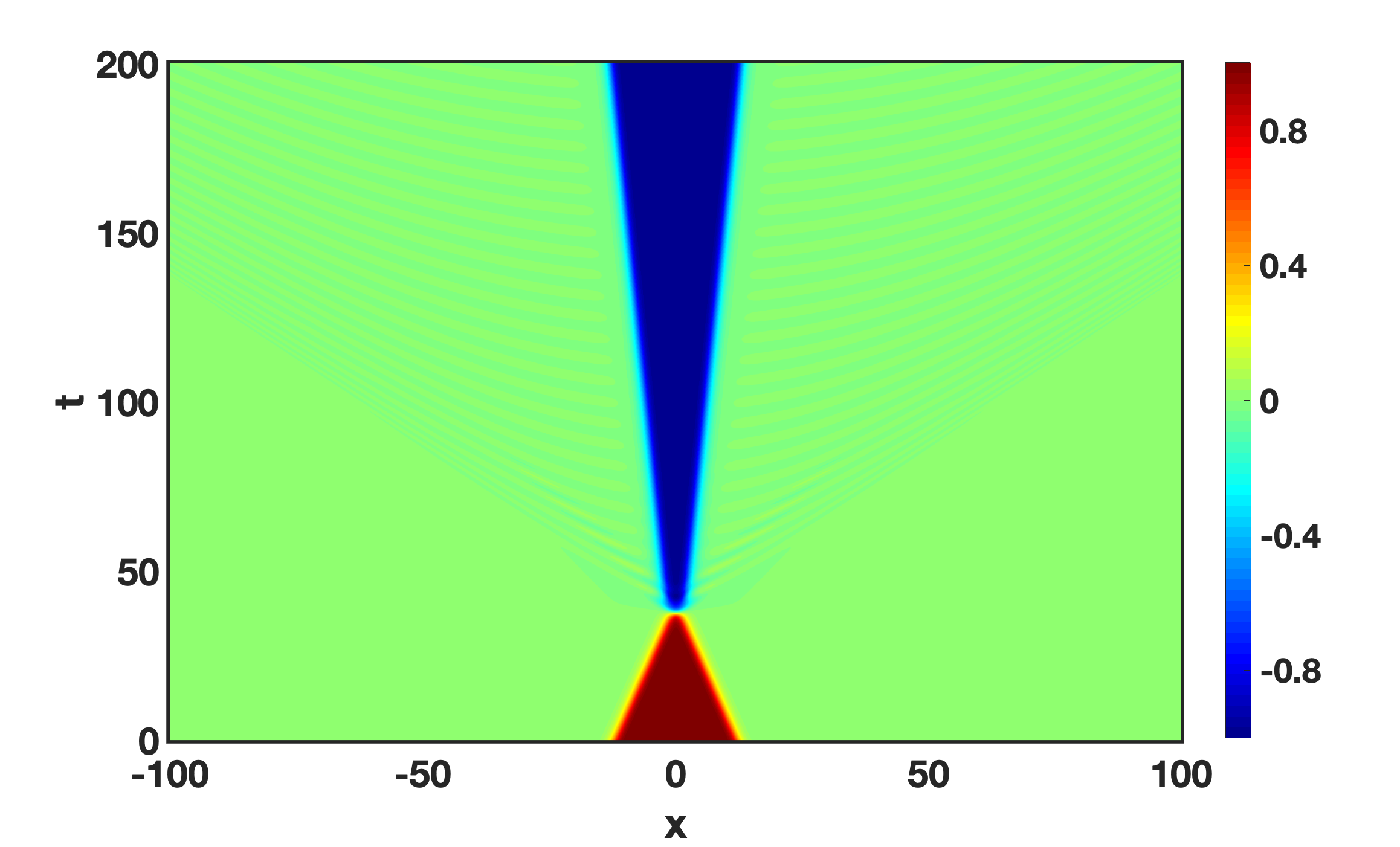}
	\caption{Kink-antikink collisions for $n=1$ with (a) $v=0.28$ - bion and (b) $v=0.30$ - one-bounce.}
	\label{n1}
\end{figure}

We analyzed the collisions varying the initial velocity $v$ and the parameter $n$. First of all, we observe that when $n=1$ we recover the same results achieved for the kink-antikink collision of the $\phi^6$ model. In particular, for $v<v_c \approx 0.289$, the kink-antikink pair always become trapped in $x=0$, as we can see in the Fig. \ref{n1}a. On the other hand, with $v>v_c$, the $K \bar K$ pair escapes to the other vacuum after the first impact - see the Fig. \ref{n1}b.

\begin{figure}
	\includegraphics[{angle=0,width=10cm,height=5cm}]{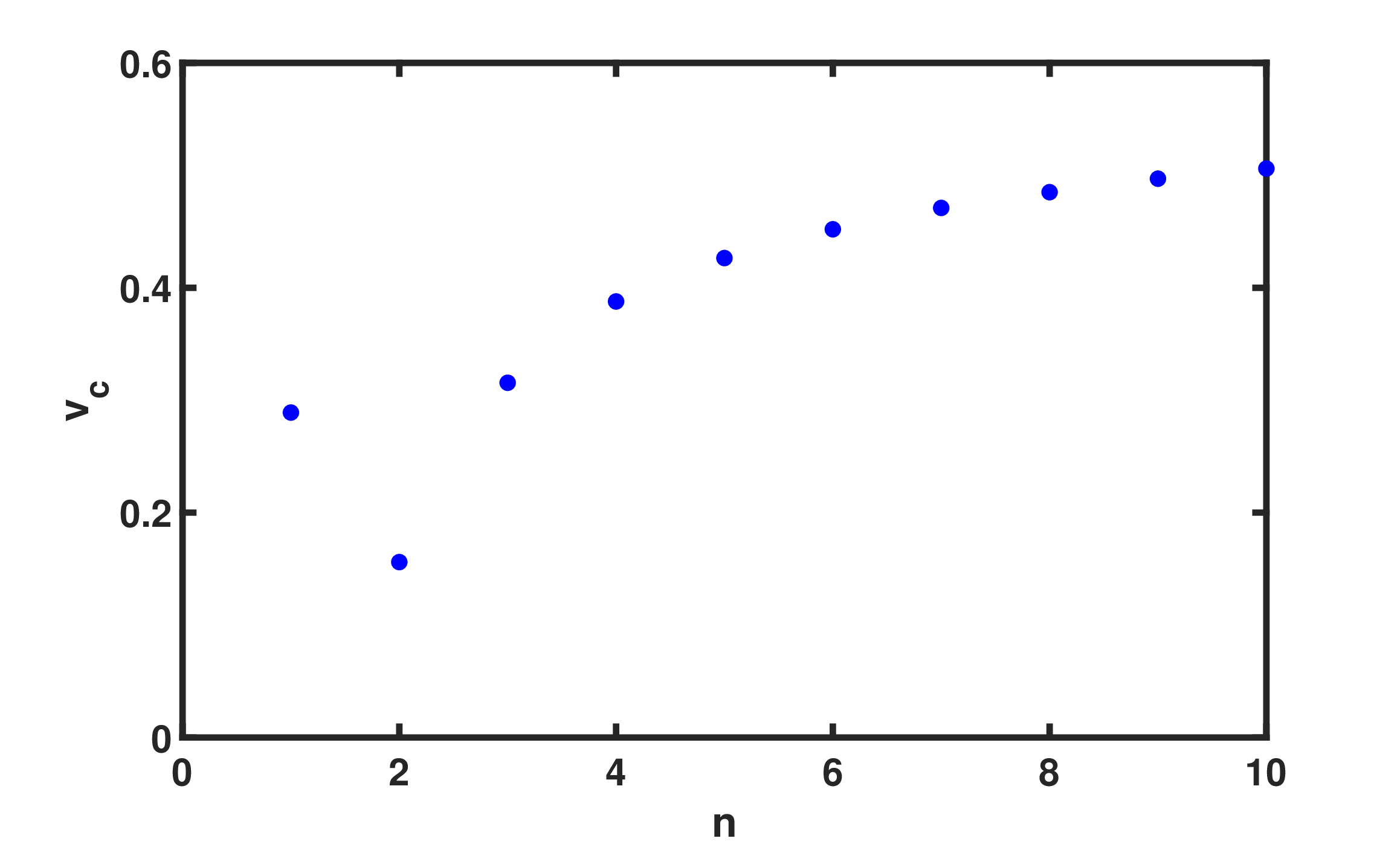}
	\caption{Kink-antikink collision. Critical velocity as a function of $n$.}
	\label{crivel}
\end{figure}

We illustrate in the Fig. \ref{crivel} the dependence of critical velocity $v_c$ versus $n$. The plot shows that $v_c$ has a minimum at $n=2$, where $v_c=0.1516$. For $n \geq 3$ we observe the enlargement of the region of bion states.

\begin{figure}
	\includegraphics[{angle=0,width=8cm,height=5cm}]{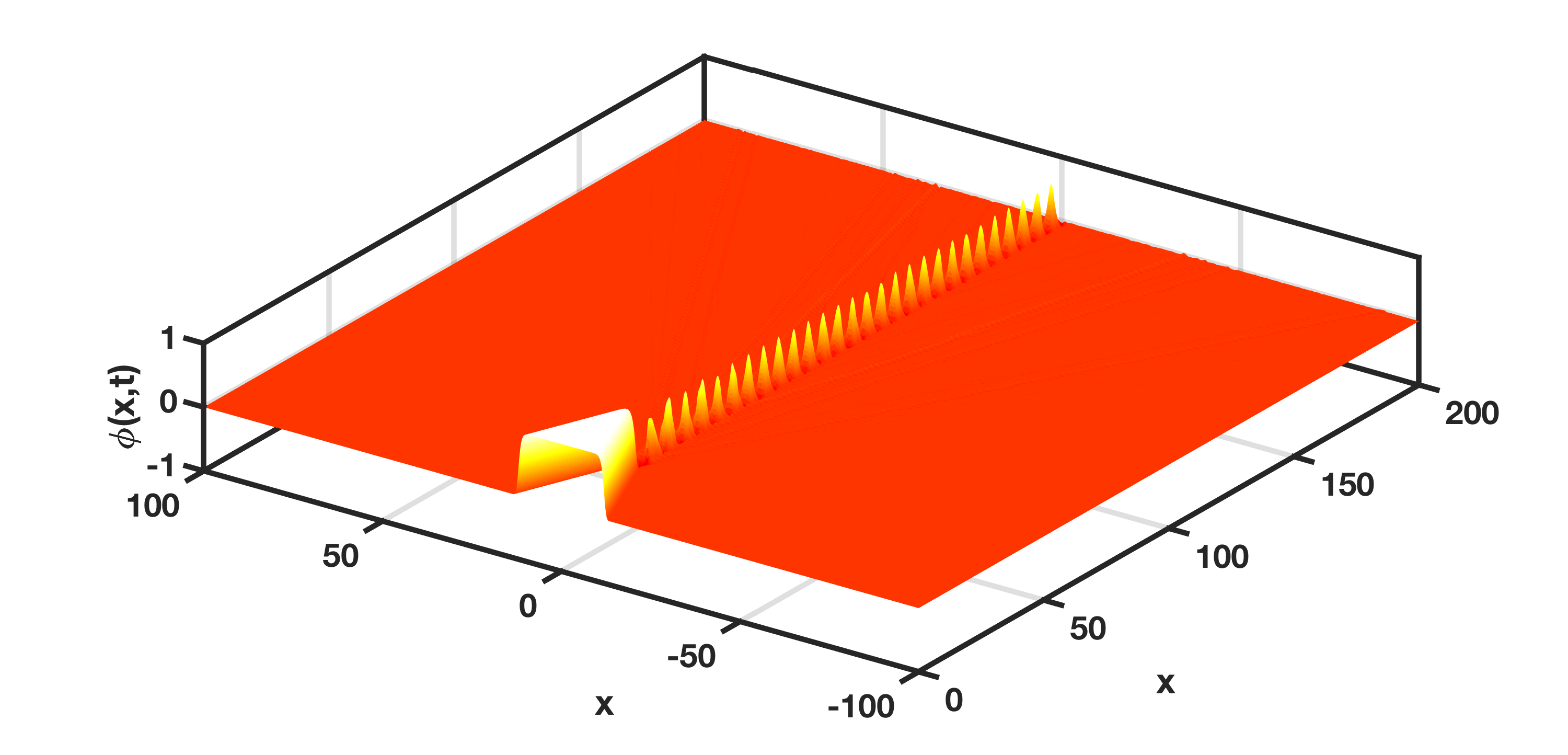}
	\includegraphics[{angle=0,width=8cm,height=5cm}]{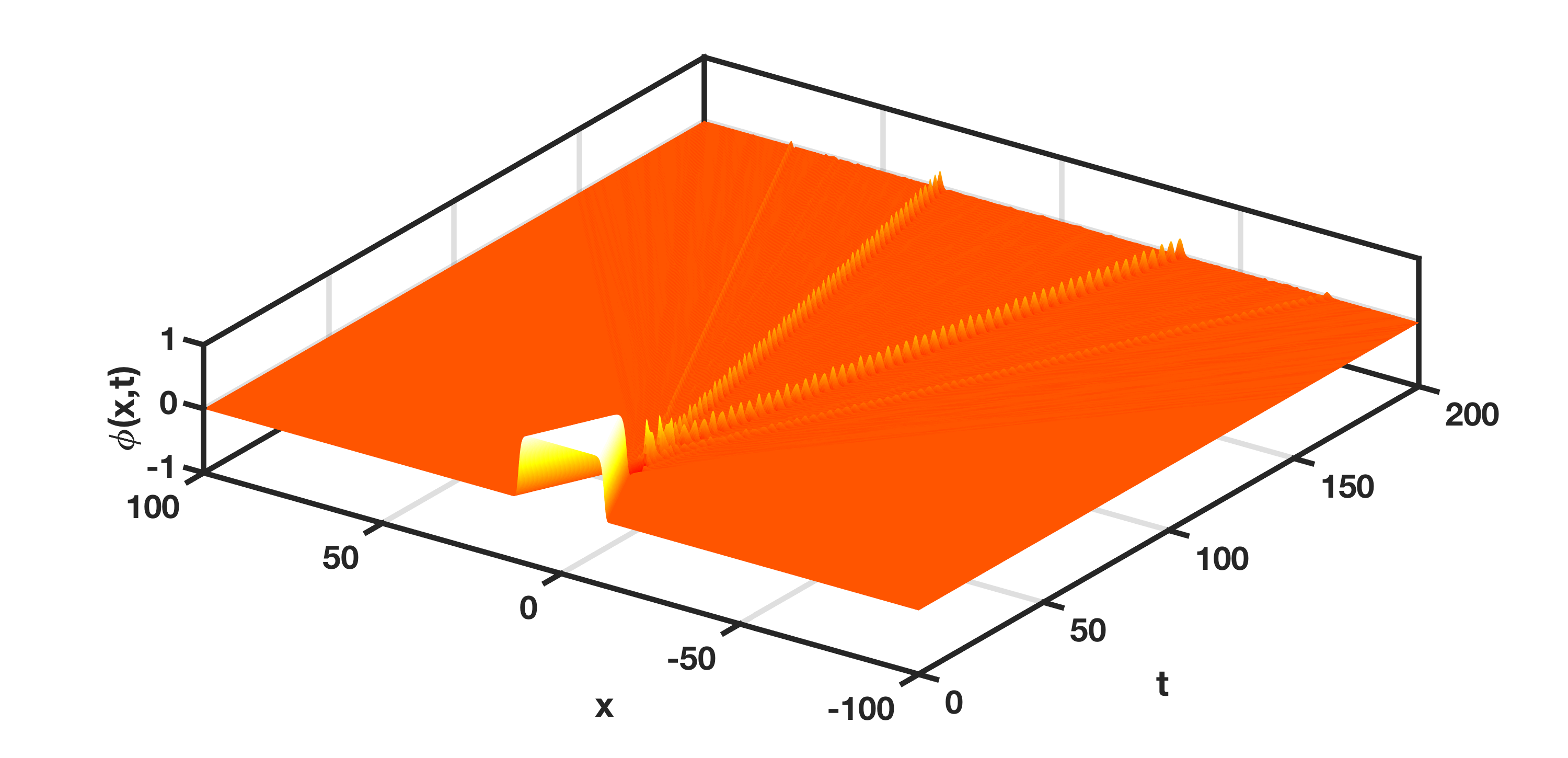}
	\includegraphics[{angle=0,width=8cm,height=5cm}]{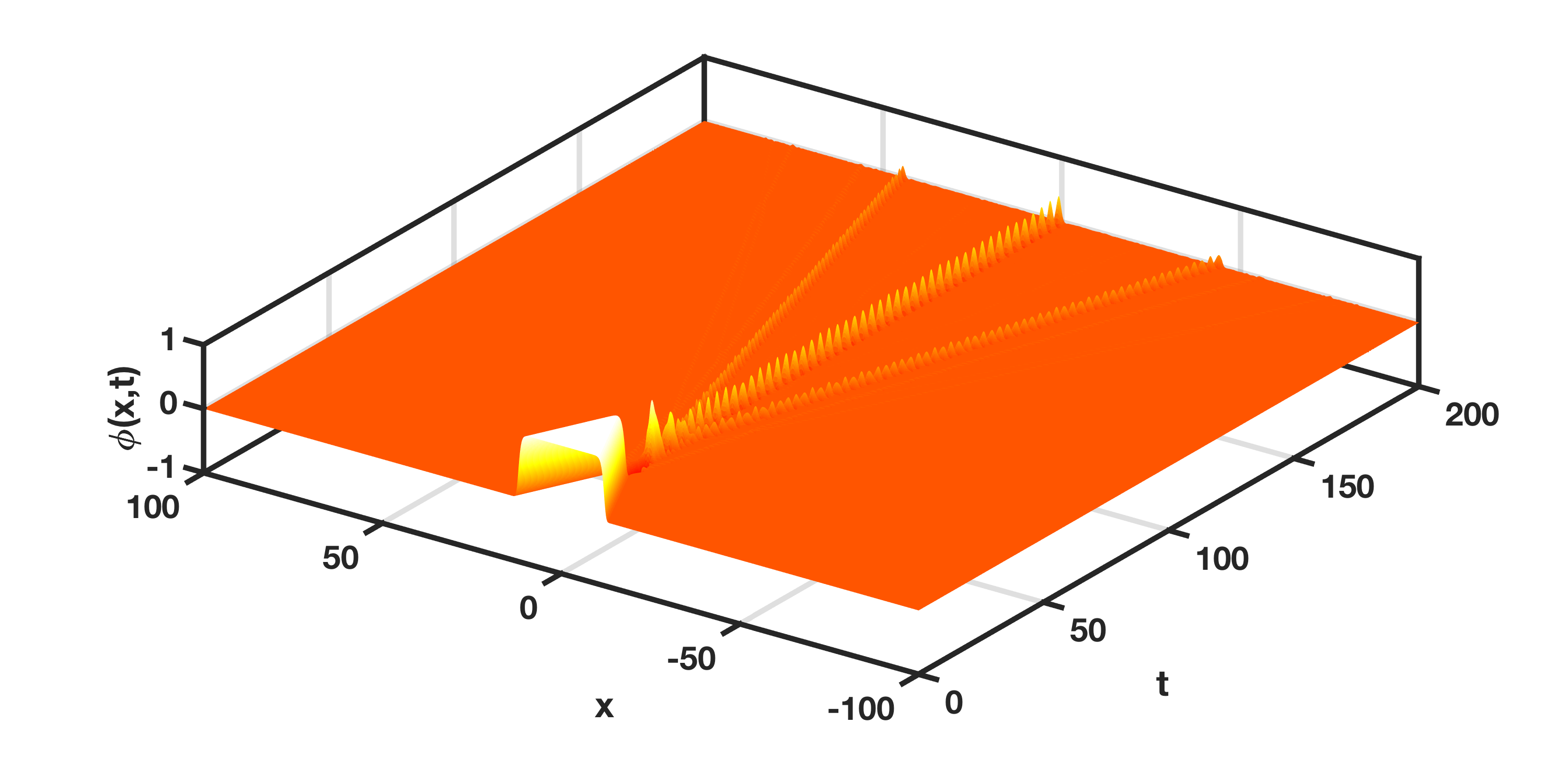}	
	\caption{Kink-antikink collisions and formation of oscillating pulses for $n=7$ with (a) $v=0.3896$ - bion, (b) $v=0.445$ - two pulses and (c) $v=0.4562$ - three pulses.}
	\label{pulse}
\end{figure}

\begin{figure}
	\includegraphics[{angle=0,width=12cm,height=4cm}]{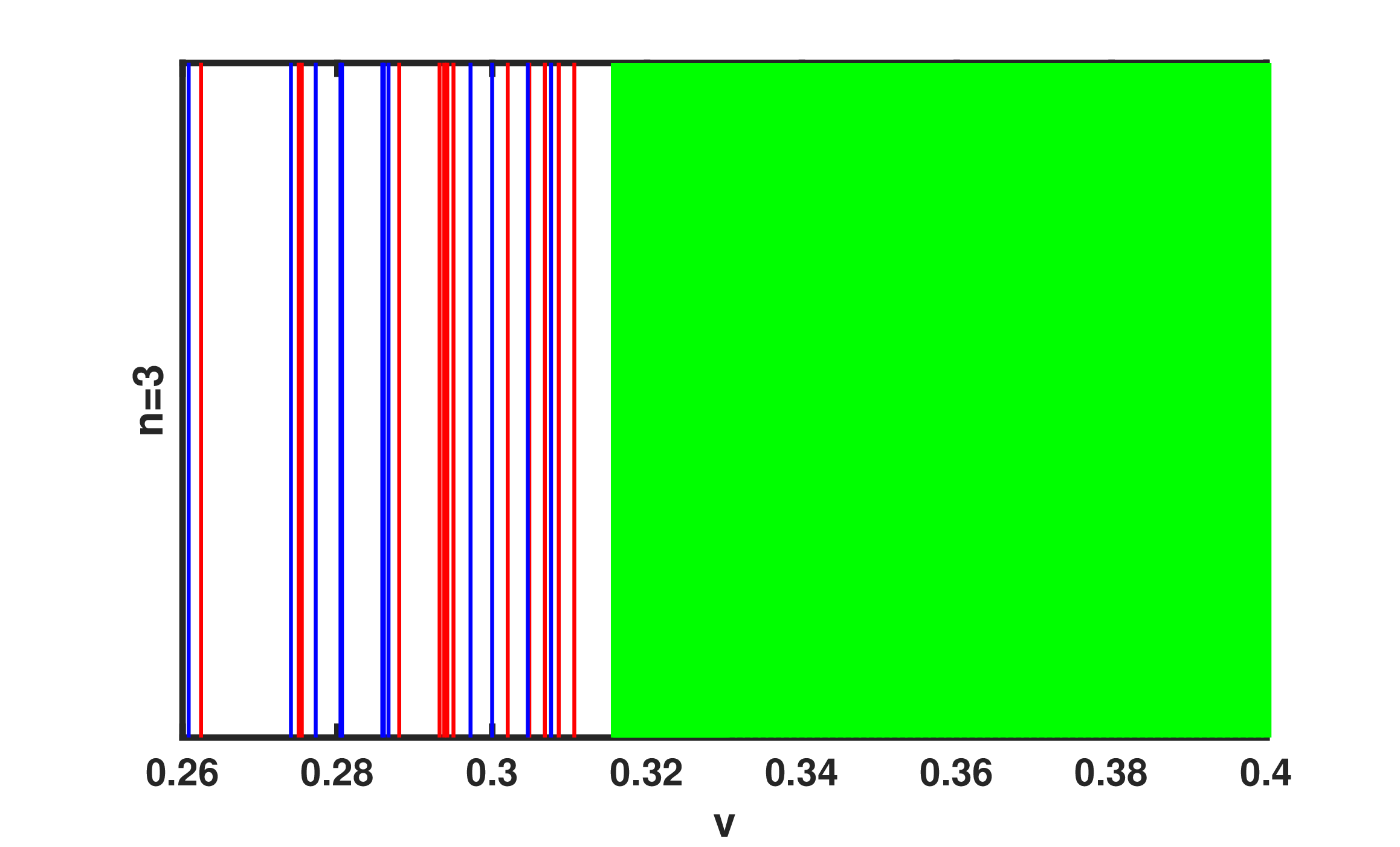}\\
	\includegraphics[{angle=0,width=12cm,height=4cm}]{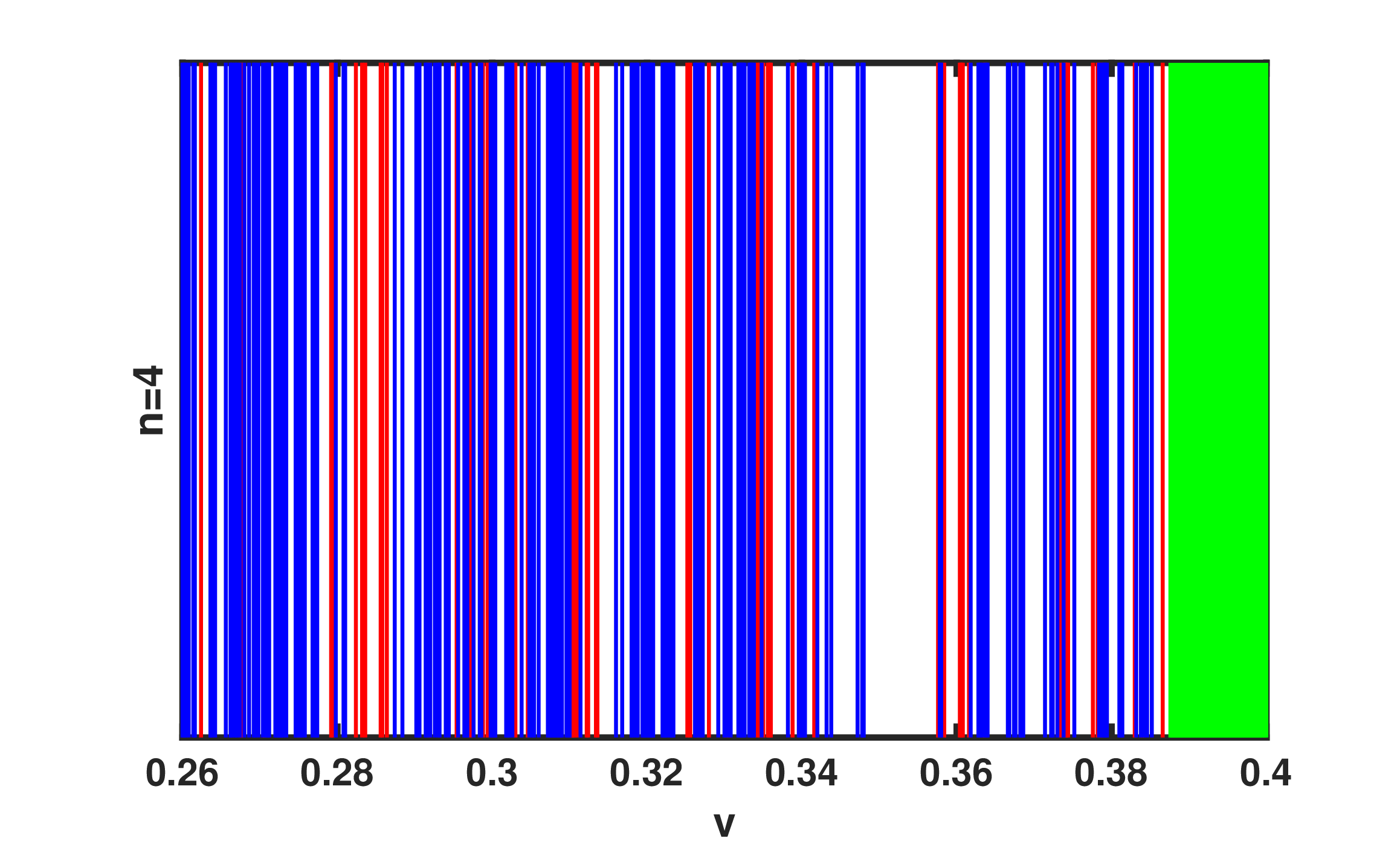}
	\caption{Kink-antikink collisions: bion state (white), two oscillations (red), three oscillations (blue) and the $K \bar K$ escapes to the other vacuum (green) for (a) $n=3$ and (b) $n=4$.}
	\label{oscka1}
\end{figure}

\begin{figure}
	\includegraphics[{angle=0,width=12cm,height=4cm}]{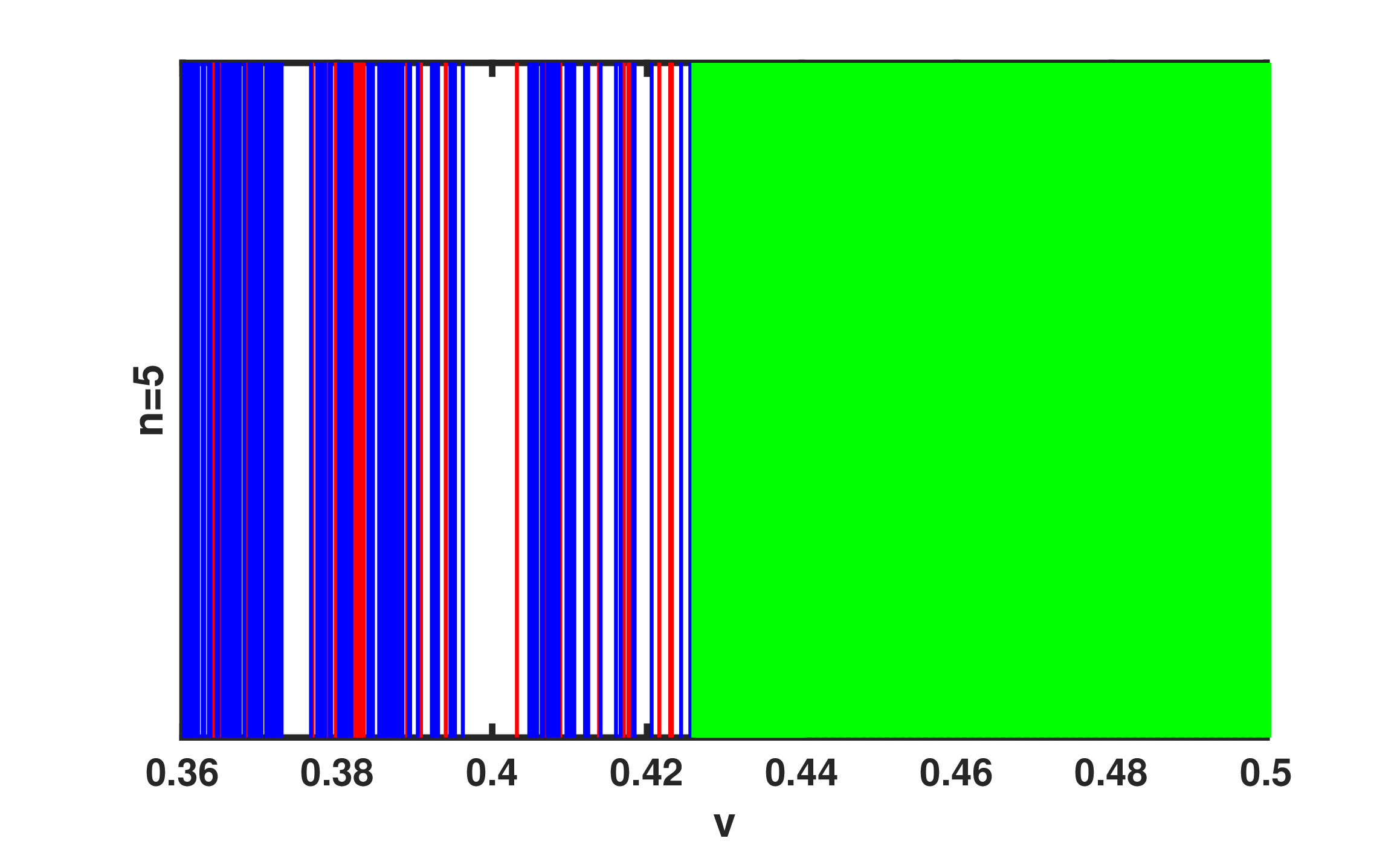}\\
	\includegraphics[{angle=0,width=12cm,height=4cm}]{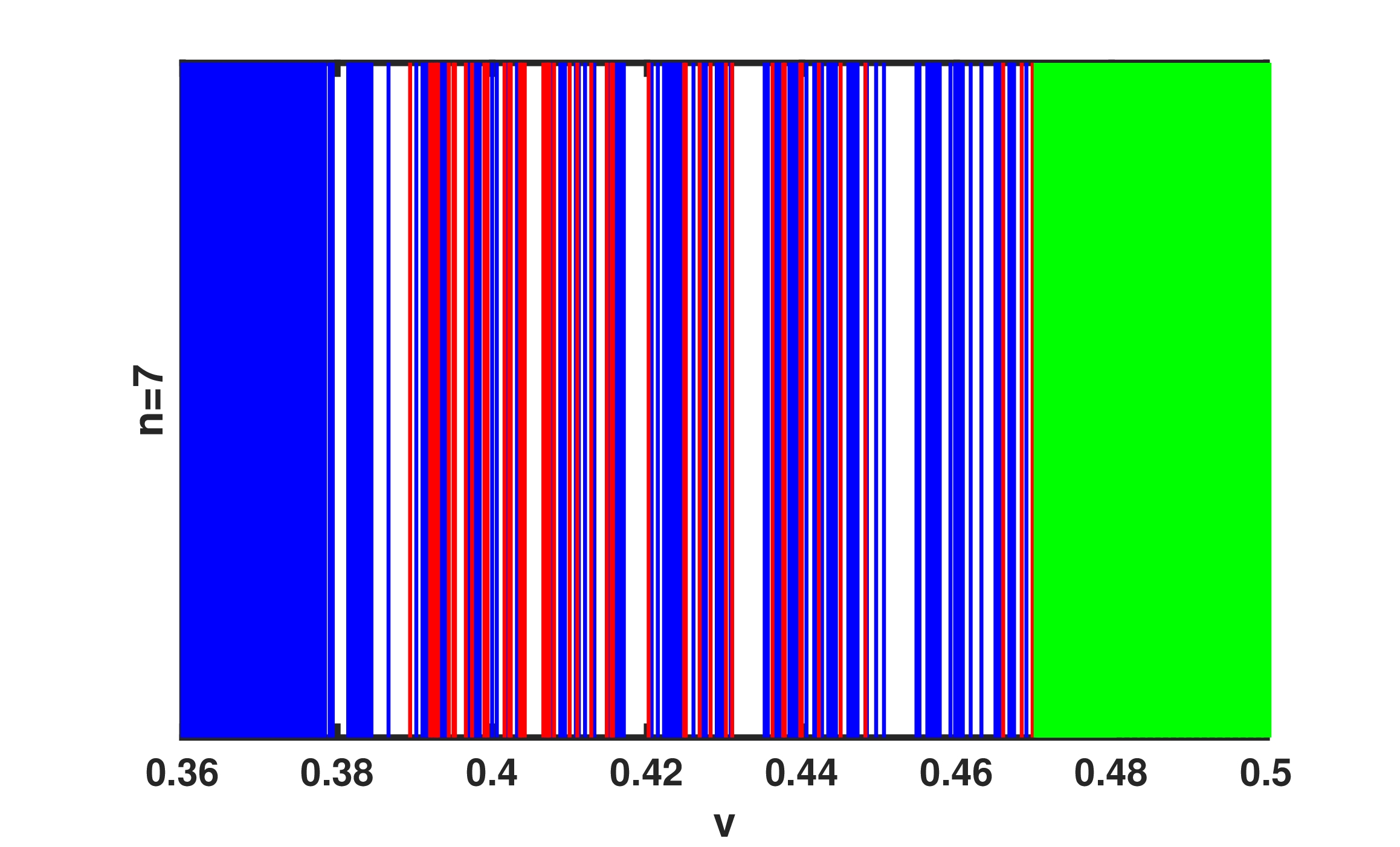}\\
	\includegraphics[{angle=0,width=12cm,height=4cm}]{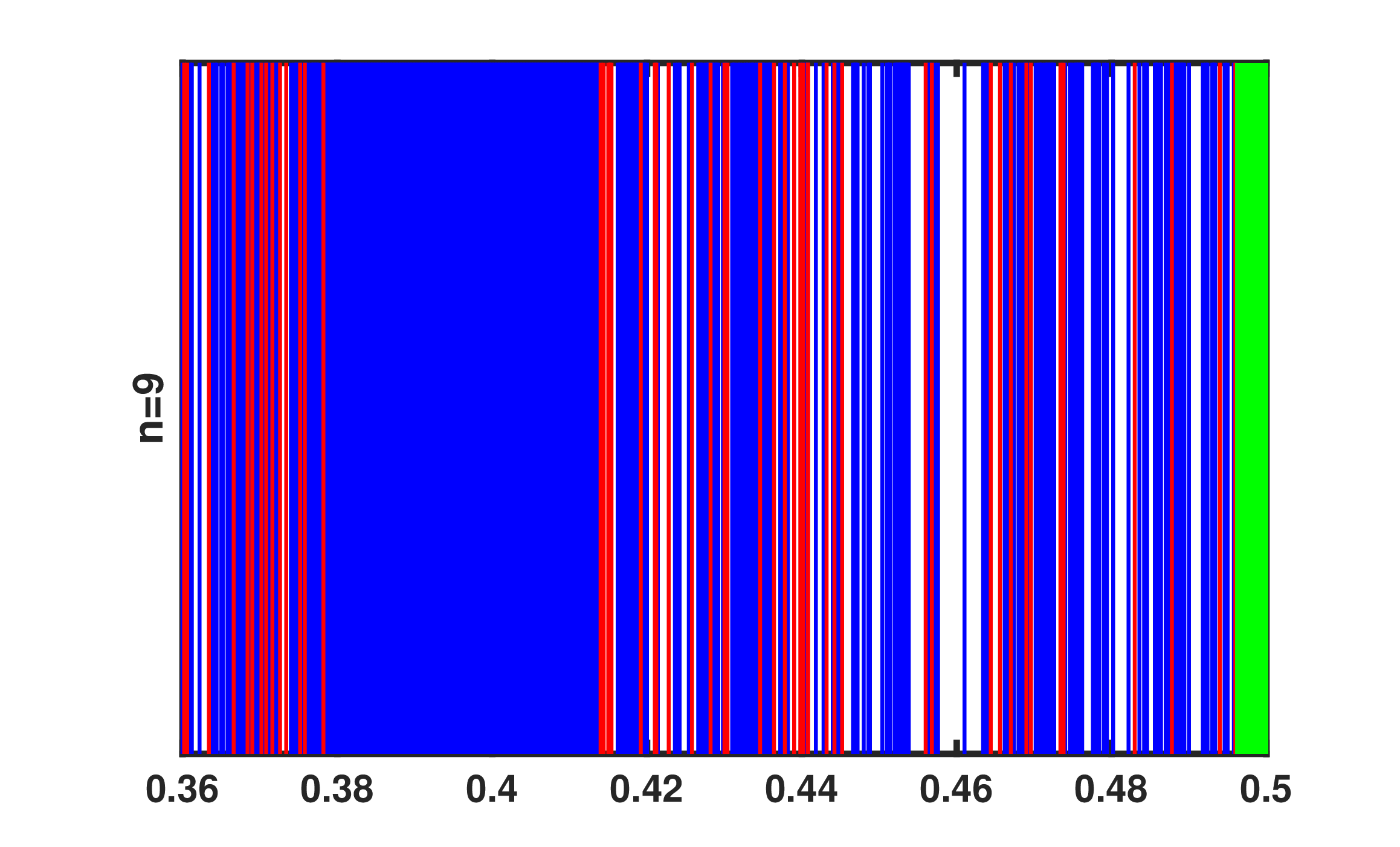}
	\caption{Kink-antikink collisions: bion state (white), two oscillations (red), three oscillations (blue) and the $K \bar K$ escapes to the other vacuum (green) for (a) $n=5$, (b) $n=7$ and (c) $n=9$.}
	\label{oscka2}
\end{figure}

Now we discuss our main results concerning the kink-antikink scattering for several values of $n \geq 2$. We report the production of oscillating pulses for some values of initial velocity and $n$. Three examples for $n=7$ are shown in the  Fig. \ref{pulse}. For instance, we observe in Fig. \ref{pulse}(a) the presence of bion state in the center of collision for $n=7$ with $v=0.3896$. On the other hand, for some ranges of initial velocity we can observe the formation of two and three oscillating pulses - see Fig. \ref{pulse}(b)-(c) for $n=7$ with $v=0.445$ and $v=0.4562$, respectively. These pulses are almost harmonic oscillations of the scalar field around one vacuum. The formation of these oscillating solutions resembles the behavior reported in Ref. \cite{roman1}, where the authors noted the presence of an oscillon in the collision of two identical wave trains. However, the oscillons could be identified when the leading frequency oscillates within the mass gap \cite{roman1}.

For small values of $n$, for example, $n=2$ these oscillations are absent. There, our results show at most the formation of a central bion. The appearance of these oscillations is sensitive to the value of the initial velocity and the $n$ parameter. The increase of $n$ contributes to the appearance of an intricate structure with greater number of oscillating pulses. As we can see in Fig. \ref{vibra}, the number of vibrational modes increase with the parameter $n$. As a consequence, this new intricate structure may be connected with the extra internal states. In the Fig. \ref{oscka1} and \ref{oscka2} we summarize some results regarding the production of oscillations as a function of the initial velocity.  In both figures, we can see that the green region corresponds to $v>v_c$, where there is the inelastic scattering of the kink-antikink pair, changing of topological sector after the collision, in a behavior analogous to the described in the Fig. \ref{n1}b for $n=1$. That is, we have the production of an antikink-kink pair: $\phi_n^{(0,1)} + \phi_n^{(1,0)} \rightarrow \phi_n^{(0,-1)} + \phi_n^{(-1,0)}$. The red and blue colors indicate the scattering of two and three pulses, respectively. The oscillations appear from $n\geq 3$. We can observe that the increase of $n$ produces an increase in the number of pulses formed.

\begin{figure}
	\includegraphics[{angle=0,width=5cm}]{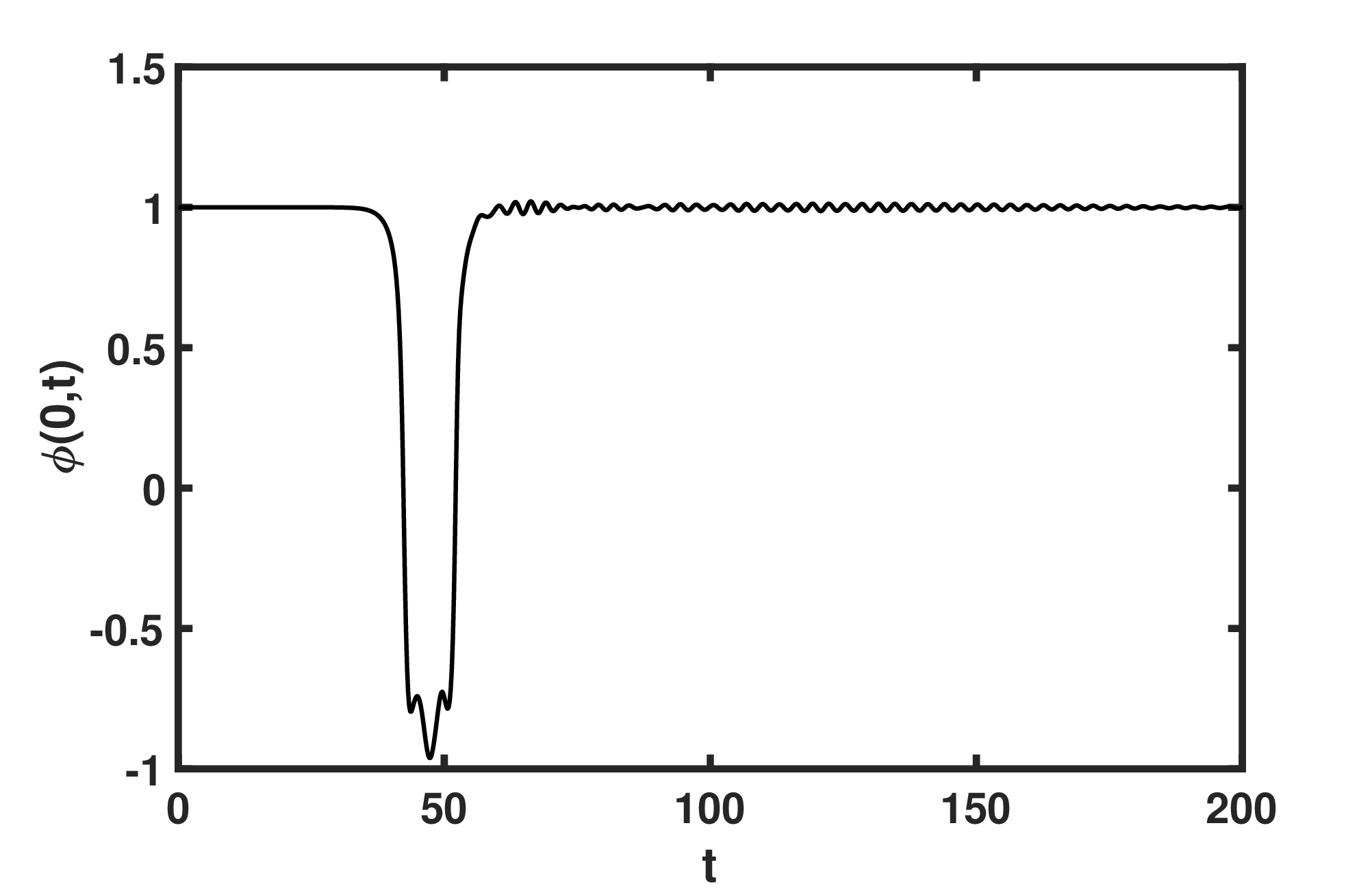}
	\includegraphics[{angle=0,width=5cm}]{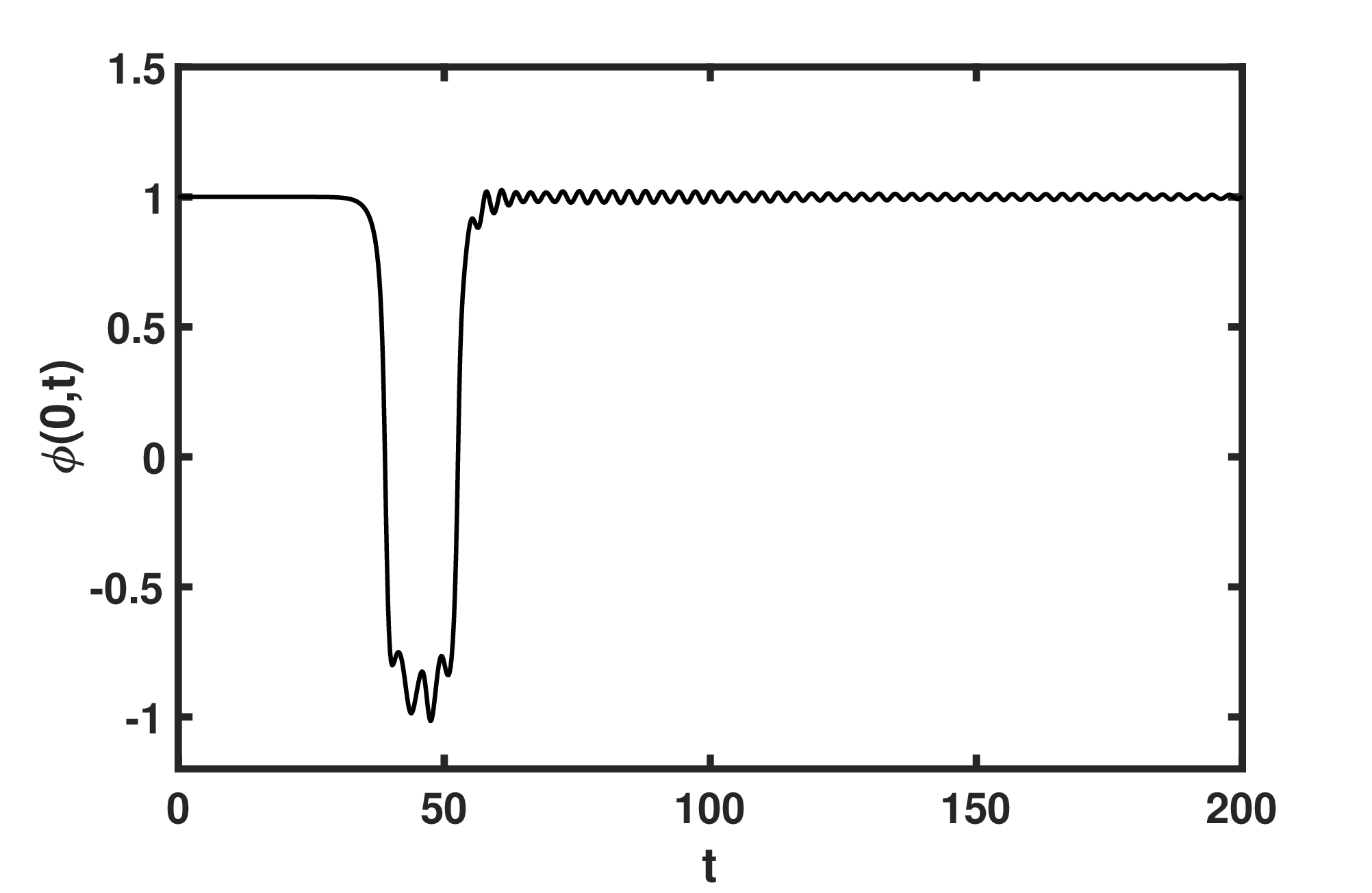}
	\includegraphics[{angle=0,width=5cm}]{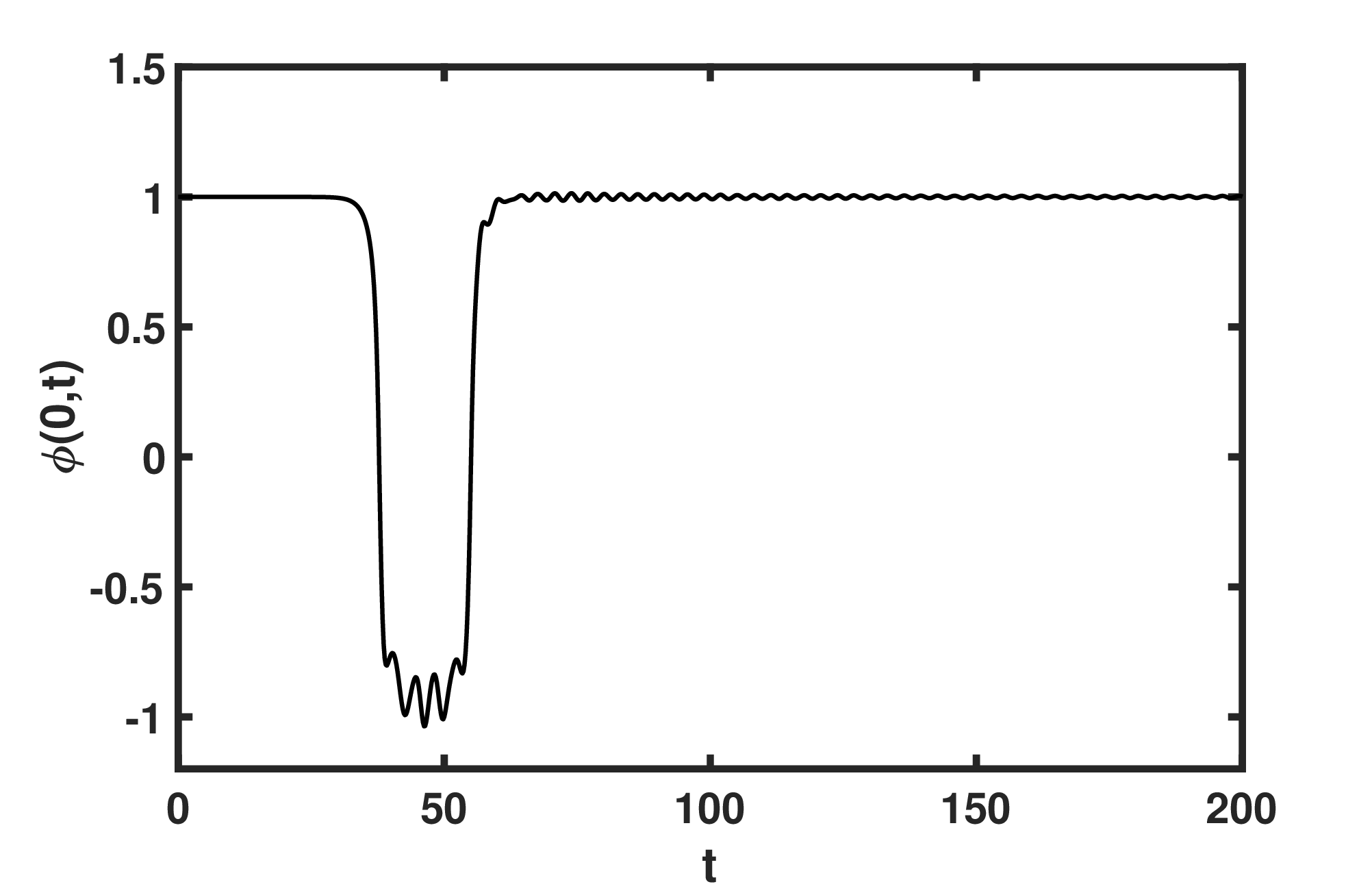}
	\caption{Kink-antikink collisions: scalar field at the center of mass $\phi(0,t)$ as a function of $t$ for (a) $v=0.265$, (b) $v=0.29$ and (c) $v=0.299$ with $N=3$, $N=4$ and $N=5$ oscillations, respectively.}
	\label{phi2d}
\end{figure}

\begin{figure}
	\includegraphics[{angle=0,width=8cm,height=5cm}]{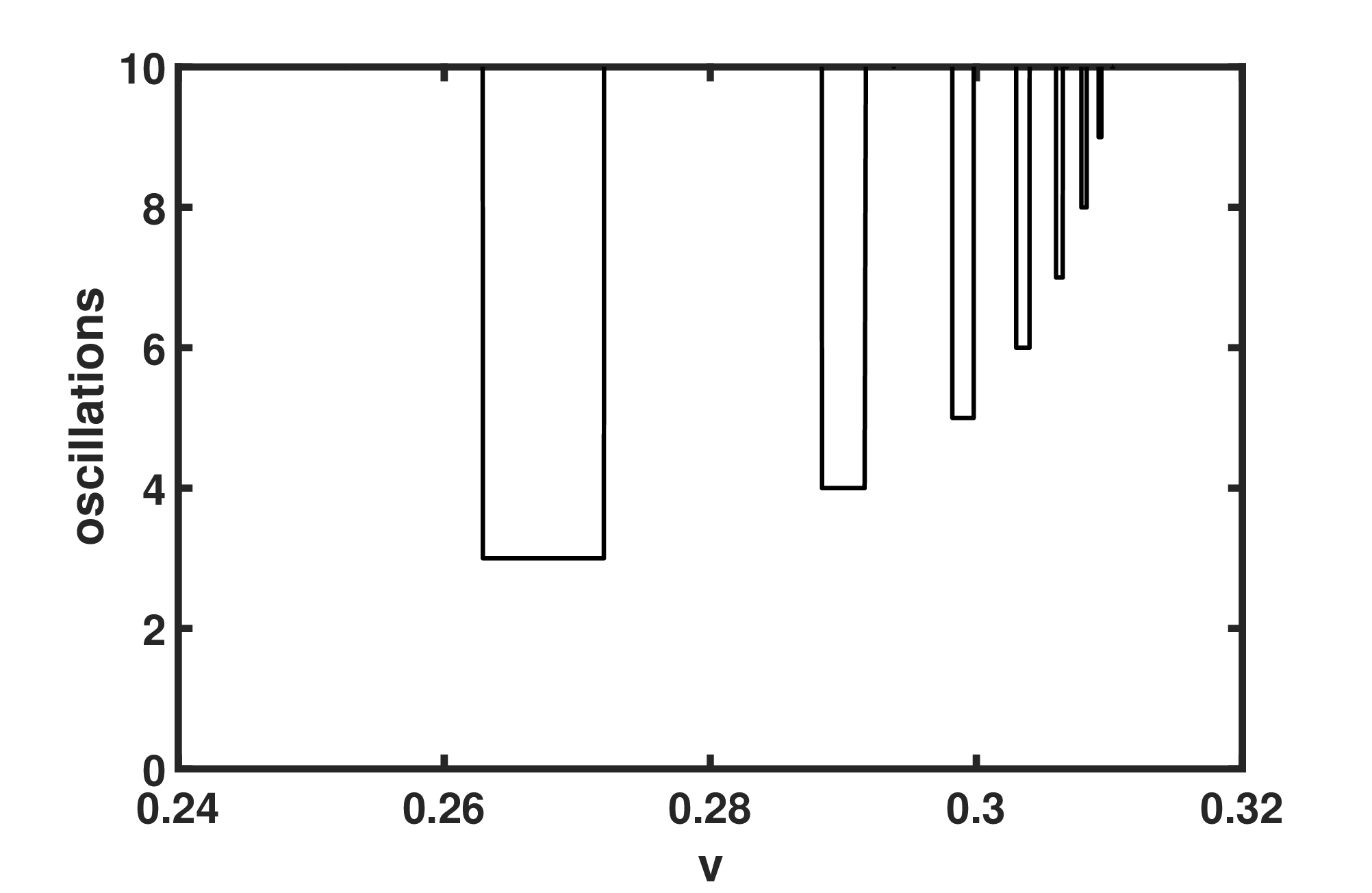}
	\caption{Number of oscillations during an one-bounce collision as a function initial velocity for $n=3$.}
	\label{osc}
\end{figure}

Another type of structure also appears for $v\lesssim v_c$: the presence of one-bounce windows. This specific type of one-bounce collision is exemplified for $n=3$ in the Figs. \ref{phi2d}a-c, which show the scalar field at the center of mass $\phi(0,t)$ as a function of $t$. There one can see that after the collision, the scalar field oscillates one (Fig. \ref{phi2d}a), two (Fig. \ref{phi2d}b) or three (Fig. \ref{phi2d}c) times around the minimum $\phi=-1$, returning to the original minimum $\phi=1$ after the bouncing. This pattern was already observed in hybrid models \cite{adalto6} and the modified sine-Gordon \cite{campbell} .

The number of oscillations during the bouncing has a structure, as shown in the  Fig. \ref{osc}. There one can see the number of oscillations in the one-bounce windows as a function of initial velocity. Note from this figure that the thickness of resonant structure decreases with $v$, when  the number of oscillations during the bouncing increases. In particular, the scattering described in the  Figs. \ref{phi2d}a-c corresponds, respectively, to the first three one-bounce windows from Fig. \ref{osc}.


\subsection { Antikink-kink scattering }


For the antikink-kink scattering in the sector $\{0,1\}$, the initial conditions are given by

\begin{eqnarray}
\phi(x,0) & = & \phi_n^{(1,0)}(x+x_0,0) + \phi_n^{(0,1)}(x-x_0,0) \\
\dot{\phi}(x,0) & = & \dot{\phi}_n^{(1,0)}(x+x_0,0) + \dot{\phi}_n^{(0,1)}(x-x_0,0).
\end{eqnarray}

\begin{figure}
	\includegraphics[{angle=0,width=8cm,height=5cm}]{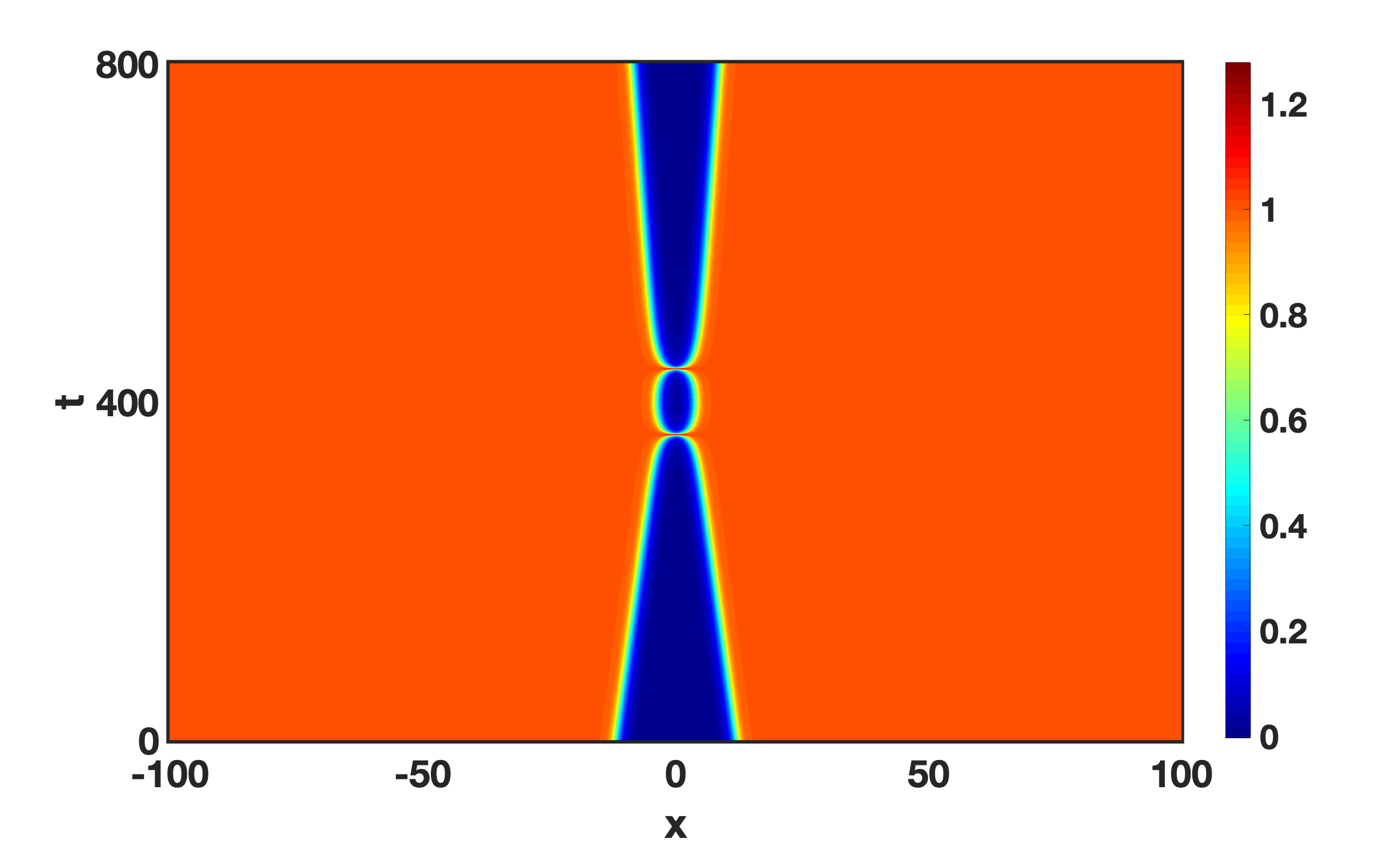}
	\includegraphics[{angle=0,width=8cm,height=5cm}]{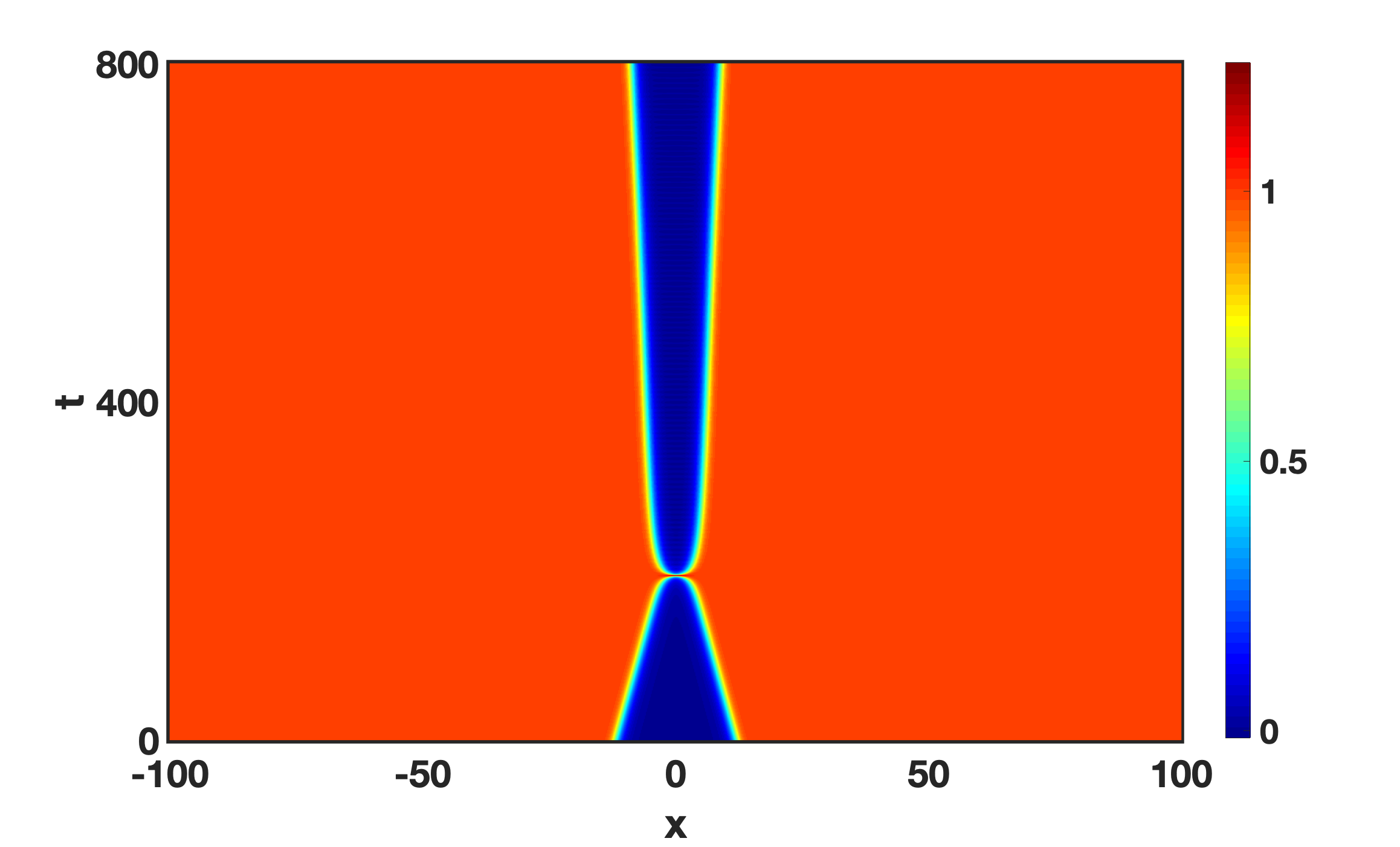}
	\caption{Antikink-kink collisions for $n=1$ with (a) $v=0.0228$ - two-bounce and (b) $v=0.046$ - one-bounce.}
	\label{AKn1}
\end{figure}

For $n=1$ we recover the known $\phi^6$ results \cite{dorey1}, with the formation of resonance structure even without the vibrational mode for a single kink or antikink for $v<v_c \approx 0.0457$ (see, for example, see Fig. \ref{AKn1}a). The reason for this was presented in Ref. \cite{dorey1} in terms of the vibrational states of the antikink-kink pair. For $v>v_c$ there is an inelastic scattering of the antikink-kink, as described for instance in the Fig. \ref{AKn1}b.

For $n=2$, we also note the appearance of the resonant window structure. For $v>v_c \approx 0.2598$, the output is an inelastic scattering between the antikink-kink pair. On the other hand, when $v<v_c$ usually occurs the production of bion states, where the scalar field $\phi(0,t)$ oscillates erratically, decaying in the long run to the minimum $\phi=1$. Furthermore, for some narrow intervals of velocity there is formation of two-bounce windows. This process is related to a resonant energy exchange between the translational and vibrational modes \cite{csw}.

It is interesting to note that the critical velocity observed for $n=2$ is very close to the critical velocity in collisions of kinks in the $\phi^4$ model. Also the squared frequency $\omega^2=3$ from Fig. \ref{fig3} is the same from the $\phi^4$ model. Indeed, the transformation $\phi \to \phi -1$ turns the model $n=2$ to the $\phi^4$ model. However, despite the symmetry of the kink, the order in which the kink-antikink or antikink-kink pair appears in the line shows the different structures in the critical velocity. Probably, the difference is due to the behavior of the potential beyond of the interval $0<\phi<1$, which contributes to the equation of motion.

\begin{figure}
	\includegraphics[{angle=0,width=8cm,height=5cm}]{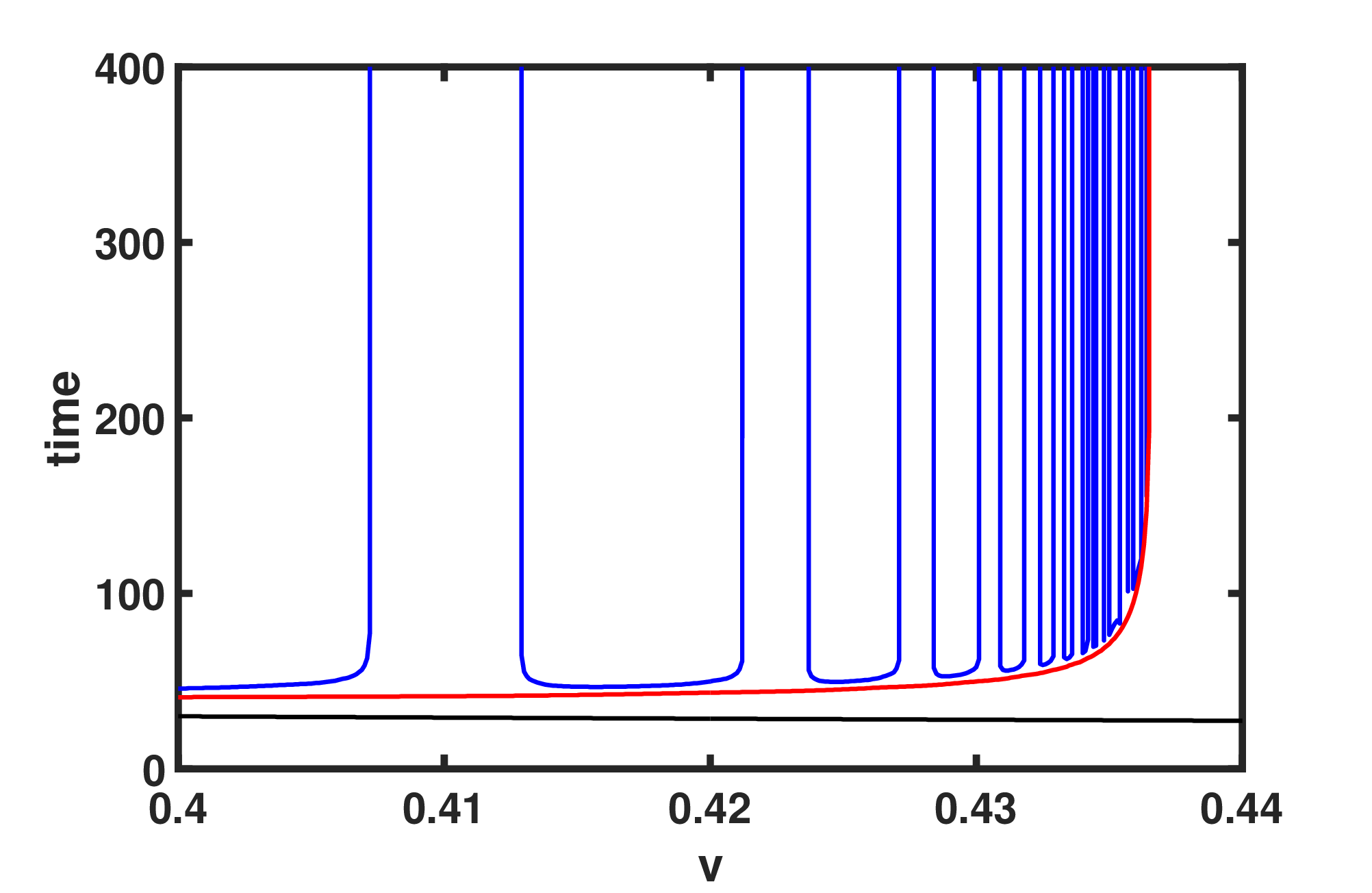}
	\includegraphics[{angle=0,width=8cm,height=5cm}]{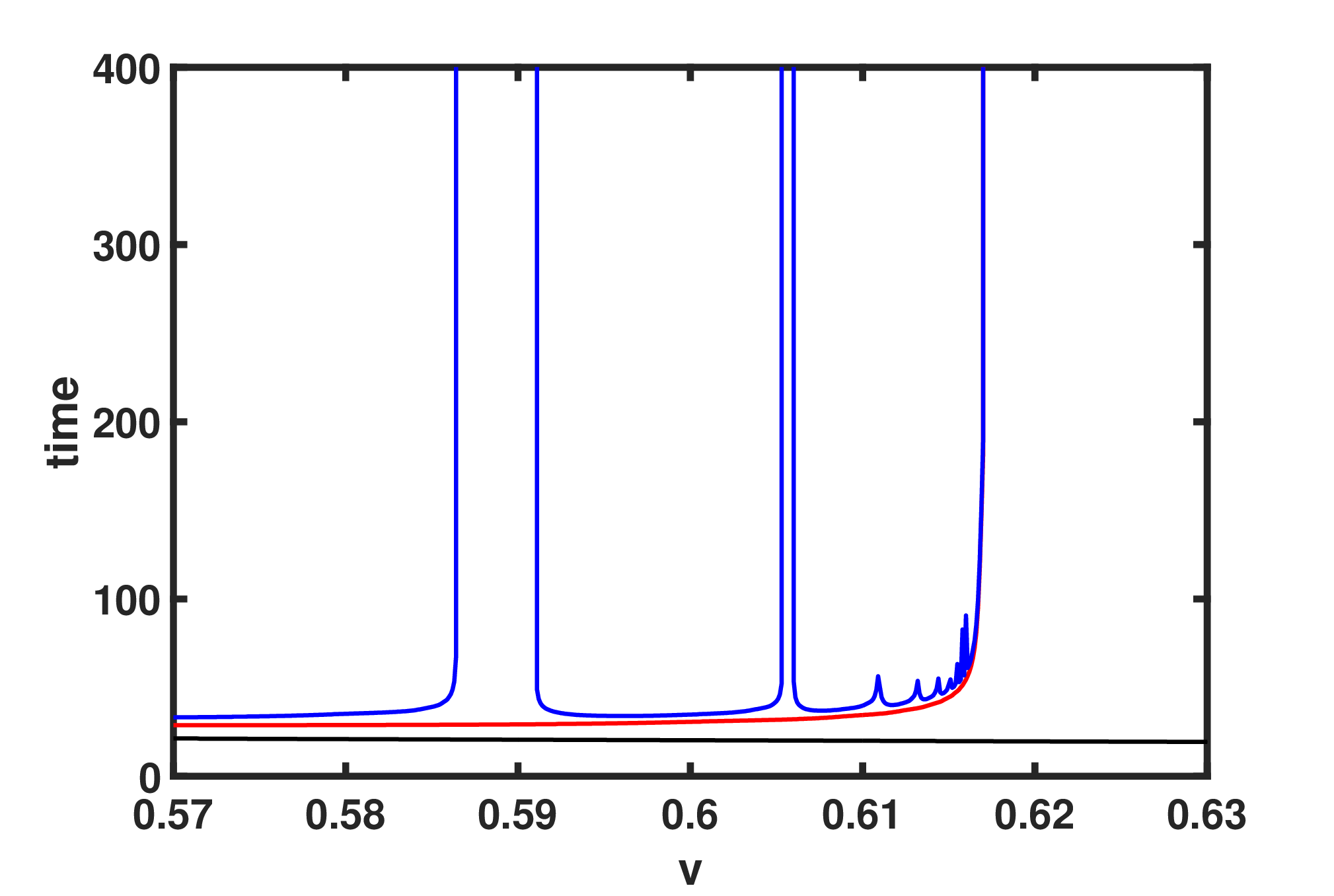}	
	\caption{Time to the first (black), second (red) and third (blue) antikink-kink collision as a function of initial velocity $v$ for (a) $n=3$ and (b) $n=5$.}
	\label{time1}
\end{figure}

\begin{figure}
	\includegraphics[{angle=0,width=5cm}]{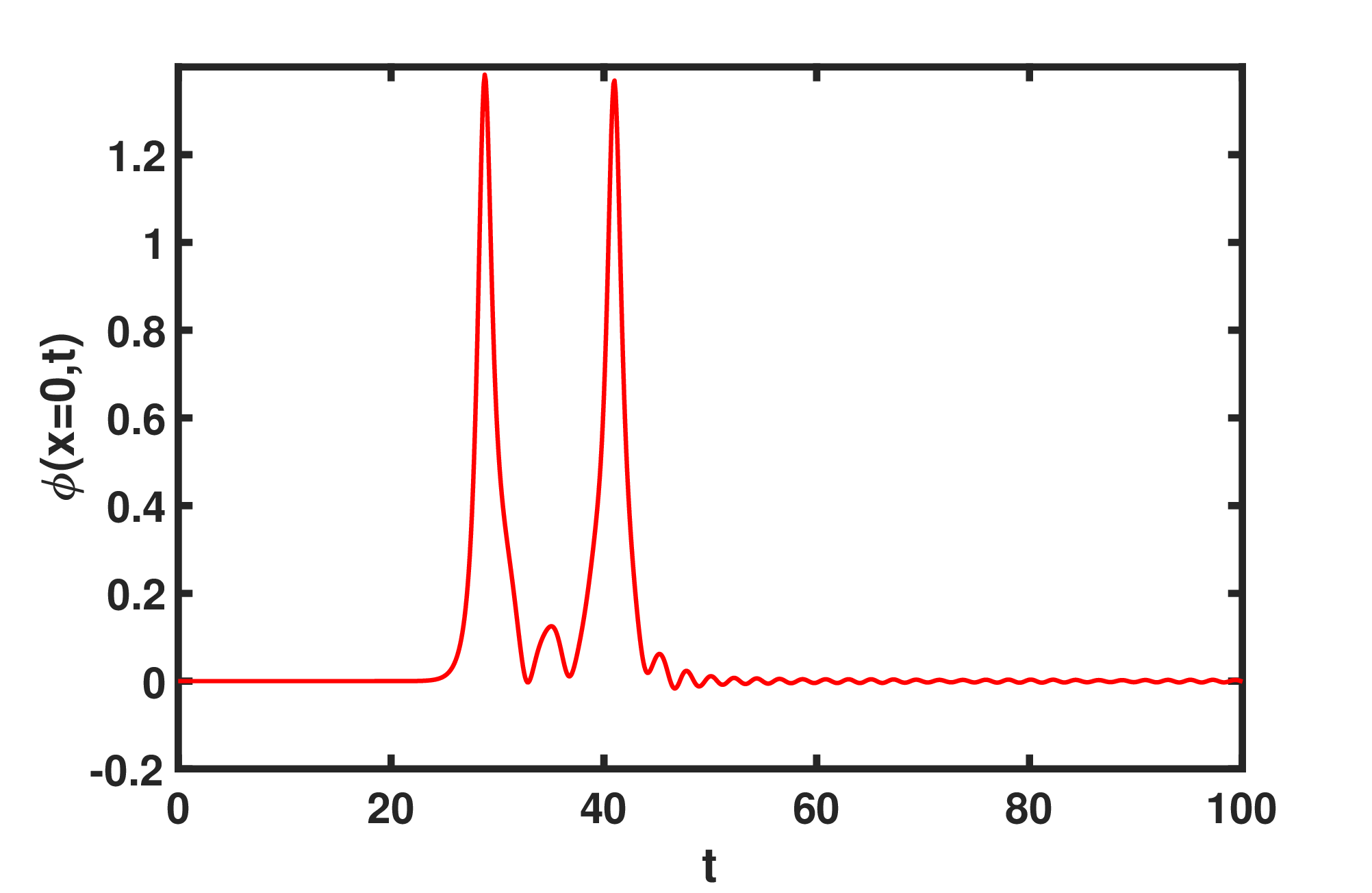}
	\includegraphics[{angle=0,width=5cm}]{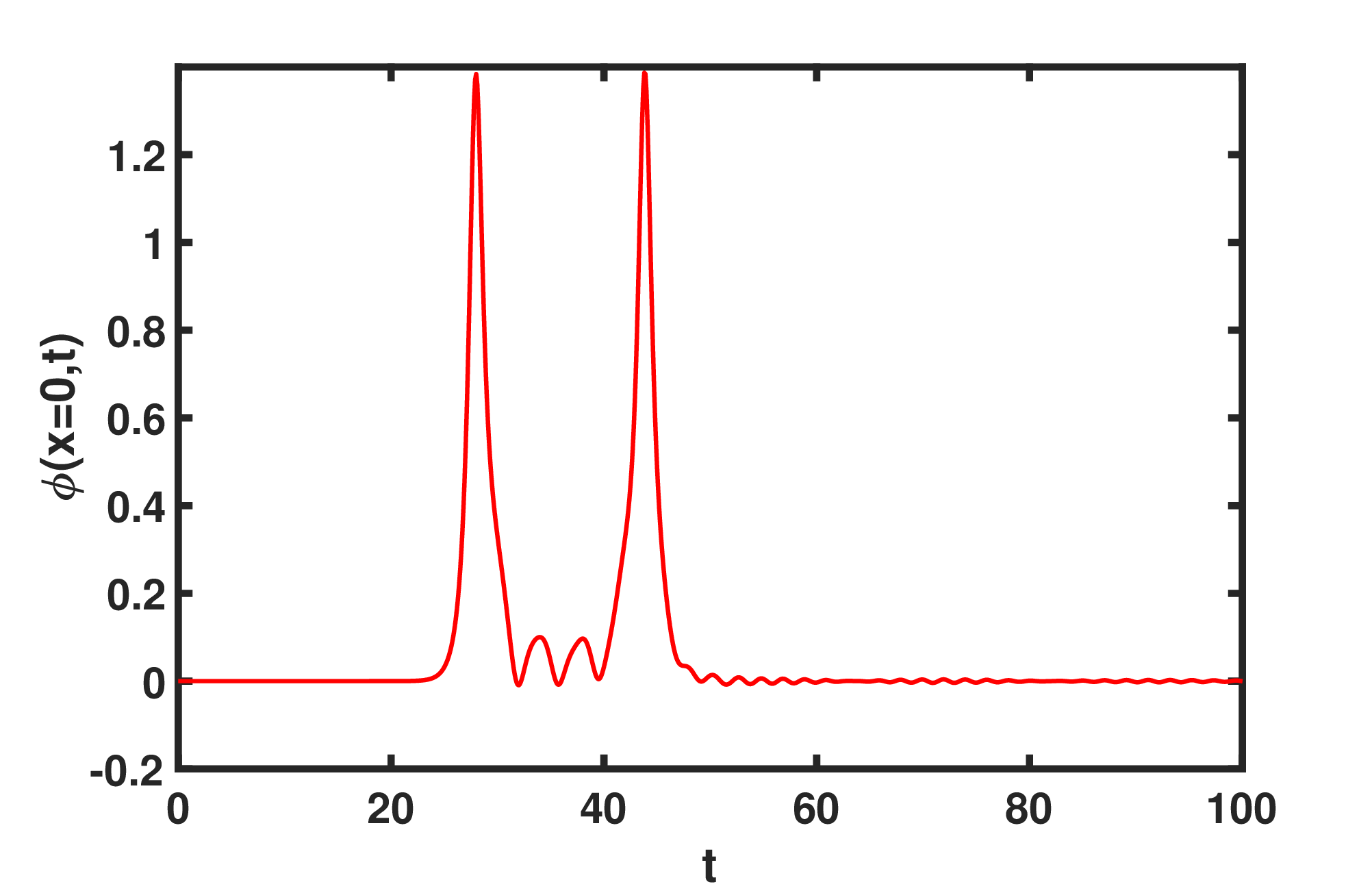}
	\includegraphics[{angle=0,width=5cm}]{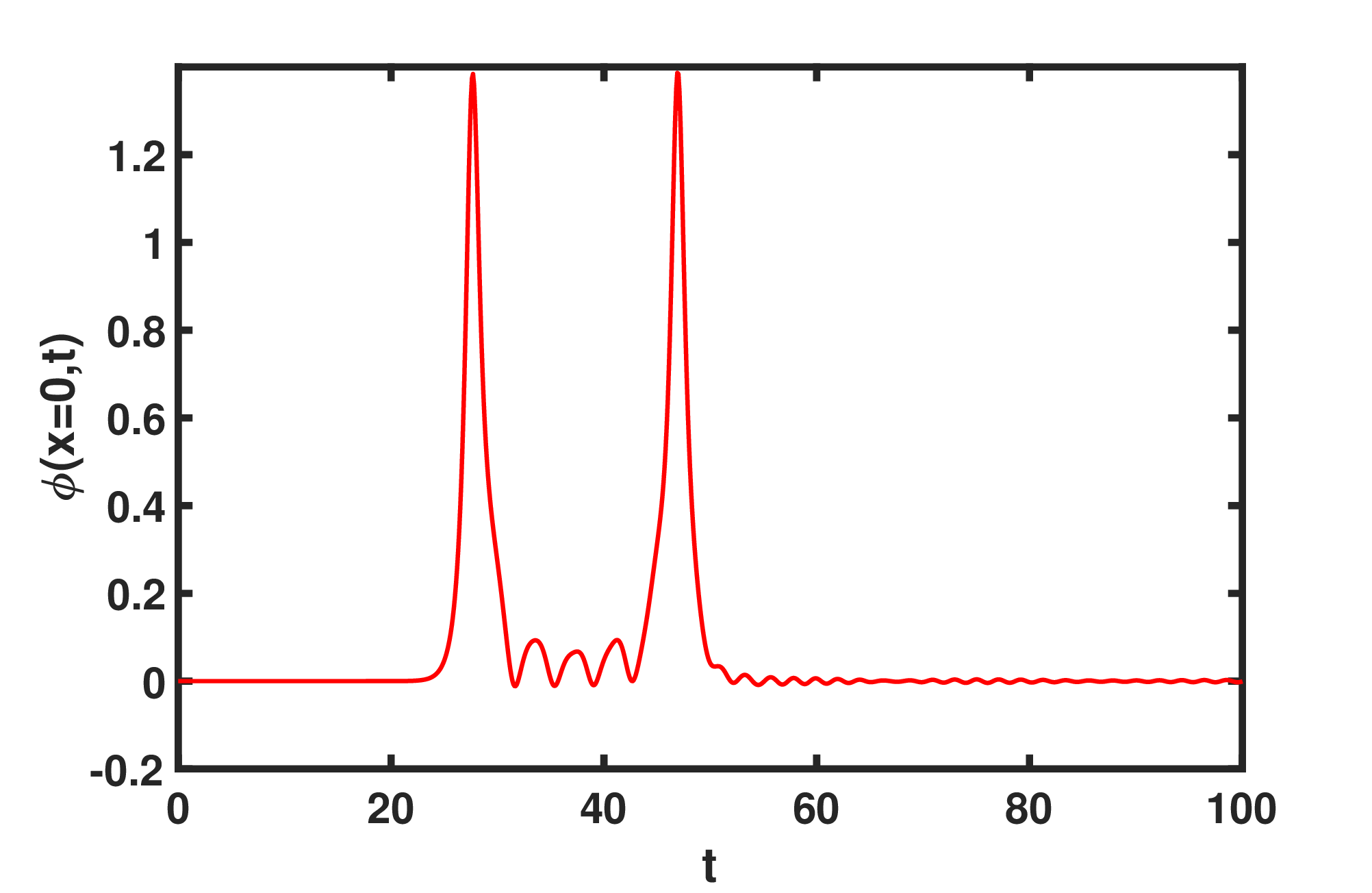}
	\caption{Antikink-kink scattering: Scalar field $\phi(x=0,t)$ at the collision center as a function of time for three consecutive two-bounce windows for $n=3$ with (a) $v=0.41$, (b) $v=0.423$ and (c) $v=0.428$.}
	\label{bounce}
\end{figure}
  
An example of the structure of two-bounce windows is presented in the Fig. \ref{time1}a, where we plot the behavior of the times of first, second and third antikink-kink collisions as a function of initial velocity of $n=3$. Note that the thickness of each two-bounce windows decreases with the velocity. For $v<v_c \approx 0.4367$, bion states or two-bounce windows occur. At the edges of each two-bounce window one sees the presence of three and four-bounce windows. Some examples of two-bounce collisions are depicted in the Figs. \ref{bounce}a-c. The two-bounce windows can be analyzed counting the number $M$ of cycle oscillations of $\phi(x=0,t)$ between each collision. Each window can then labeled by an integer $m=M-2$. For instance, Fig. \ref{bounce}a shows the plot of $\phi(x=0,t)$ oscillates $M=3$, corresponding to the expected first ($m=1$) two-bounce windows. 
In the Figs. \ref{bounce}b-c, the scalar field oscillates $M=4$ and $M=5$ times around the vacuum $\phi=0$ during the bouncing, corresponding to the second ($m=2$) and third ($m=3$) two-bounce windows, respectively.

\begin{figure}
	\includegraphics[{angle=0,width=8cm}]{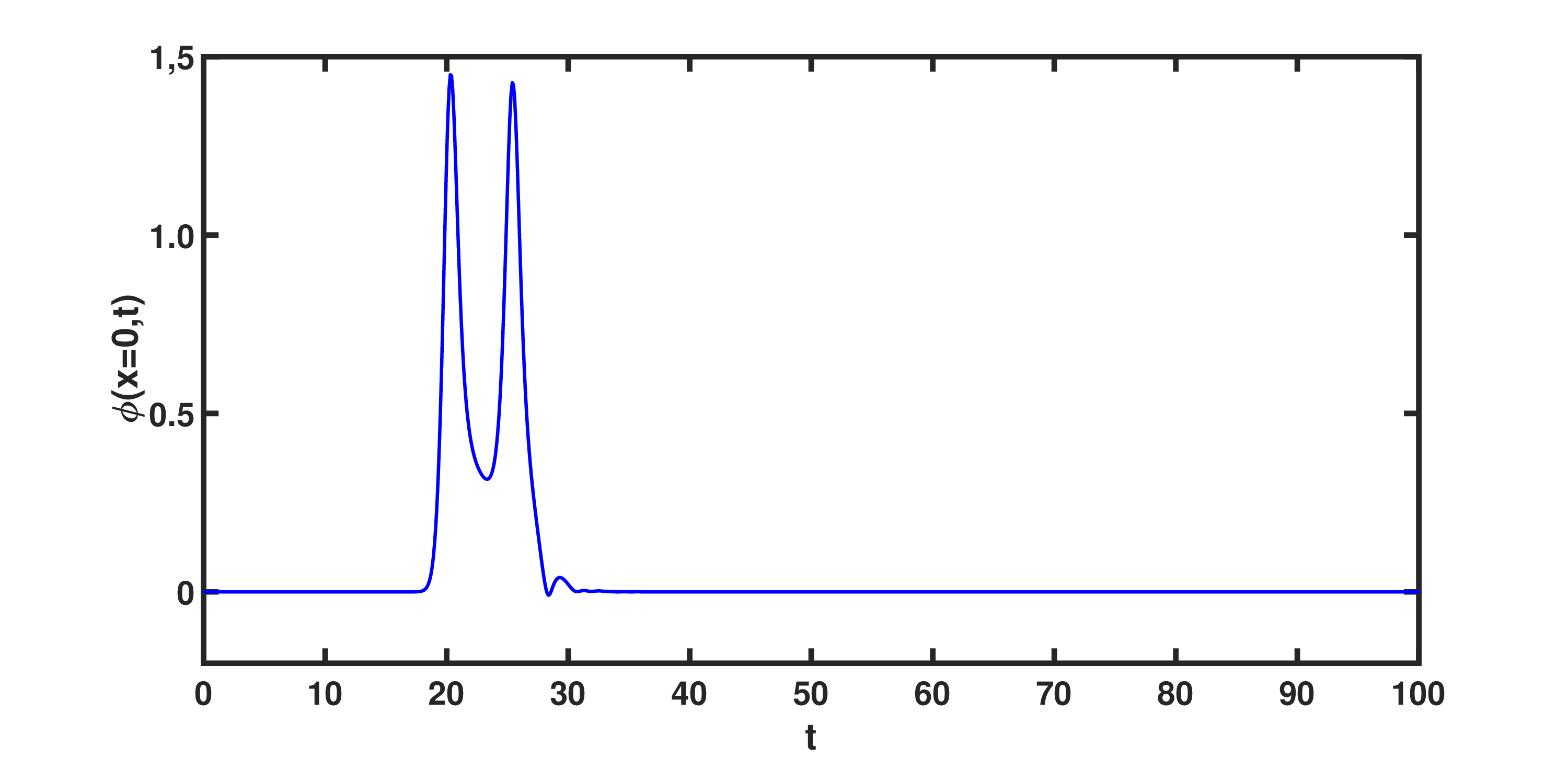} 
	\caption{Antikink-kink scattering: Scalar field $\phi(x=0,t)$ at the collision center as a function of time for $n=7$ with $v=0.6$.}
	\label{bounce1}
\end{figure}
An example of two-bounce window for a larger value of $n$ is depicted in Fig. \ref{bounce1}. In this figure we describe the behavior of $\phi(x=0,t)$ for the case $n=7$ with $v=0.60$. Note that the two bounces are characterized by two peaks. There is no oscillation between the bounces, that characterizes it as a zero-order window. The behavior is slightly different when compared to the Fig. \ref{bounce}.

 As we saw in the previous section, for $n\geq2$,  an isolated antikink or kink has always one vibrational state (see the Fig. \ref{fig3}).  The same applies for the antikink-kink pair (cf. Fig. \ref{vibra}b).
Then, one would expect the presence of two-bounce windows. However, as the value of $n$ increases, the higher order windows disappear, traded by false two-bounce windows or quasiresonances \cite{gani4}; see, for instance, the above Fig. \ref{time1}b for $n=5$. The absence of the resonance windows is related to the approximation of the value of the energy of the vibrational mode with the mass threshold - see Fig. \ref{fig3}. As a result, the translational and vibrational modes are unable to perform the resonant energy exchange mechanism.

\begin{figure}
	\includegraphics[{angle=0,width=10cm,height=6cm}]{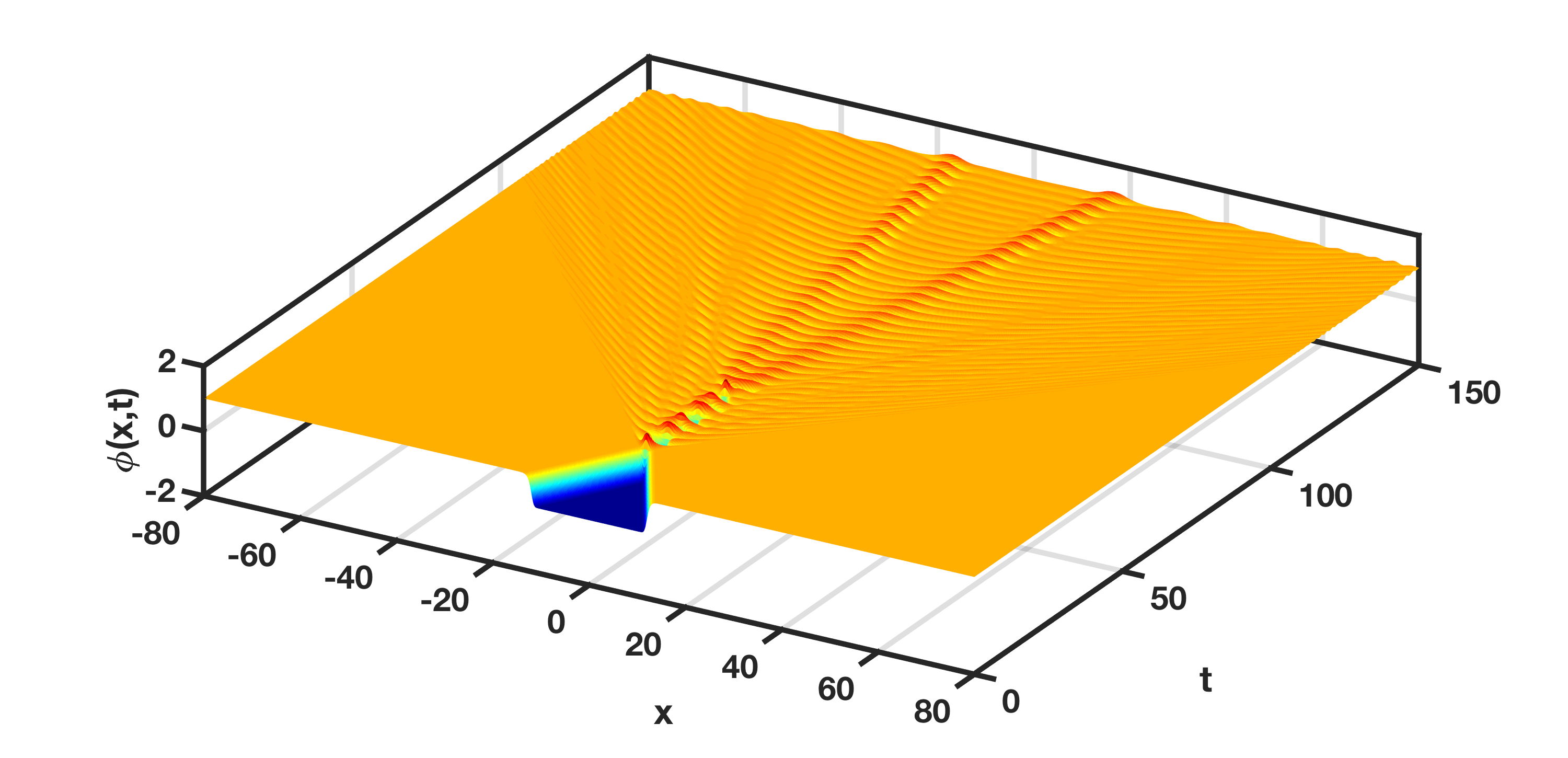}
	\caption{Antikink-kink collisions and two pulses formation for $n=6$ with $v=0.6267$.}
	\label{oscillon1}
\end{figure}

\begin{figure}
	\includegraphics[{angle=0,width=12cm,height=3cm}]{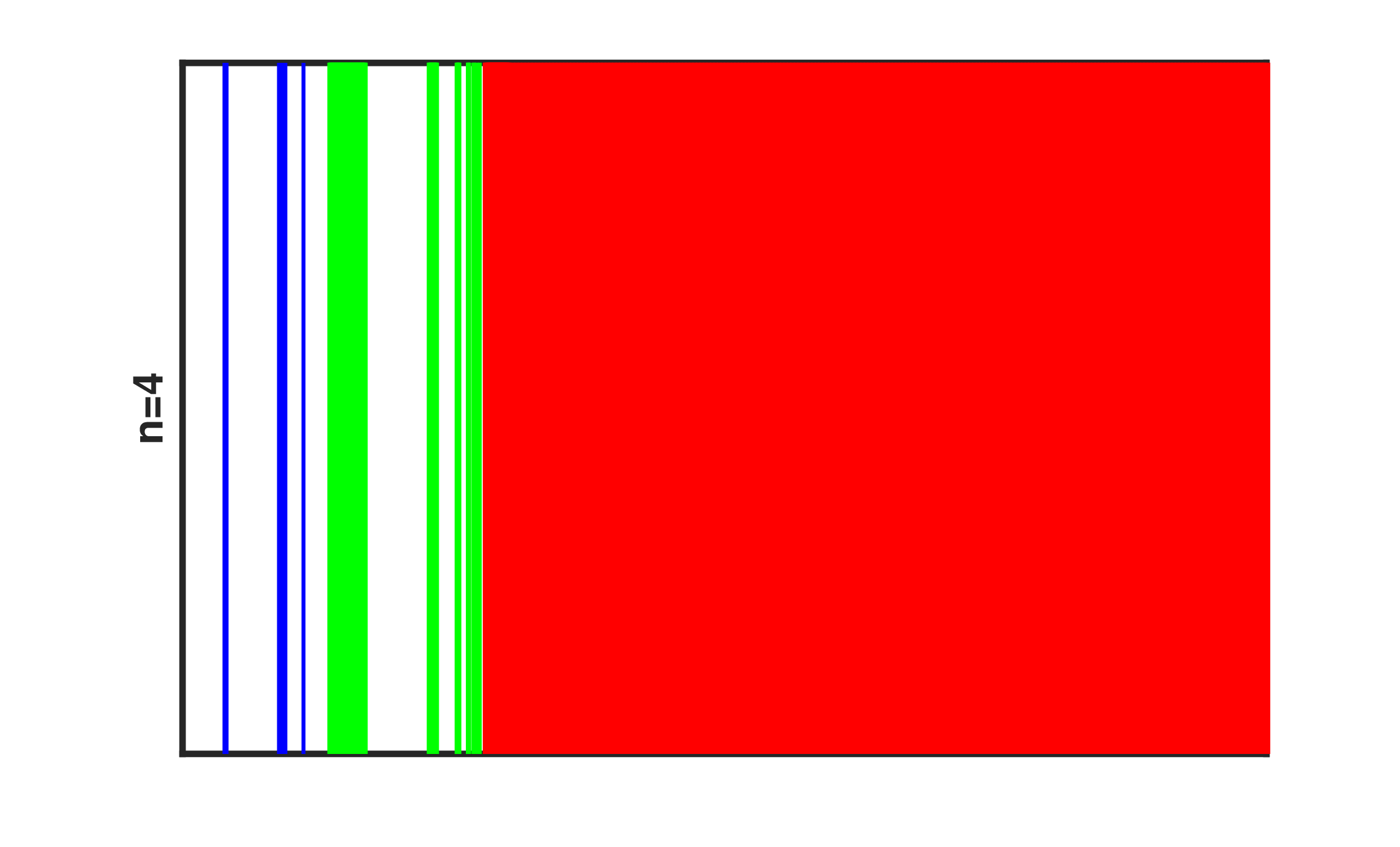}\\
	\includegraphics[{angle=0,width=12cm,height=3cm}]{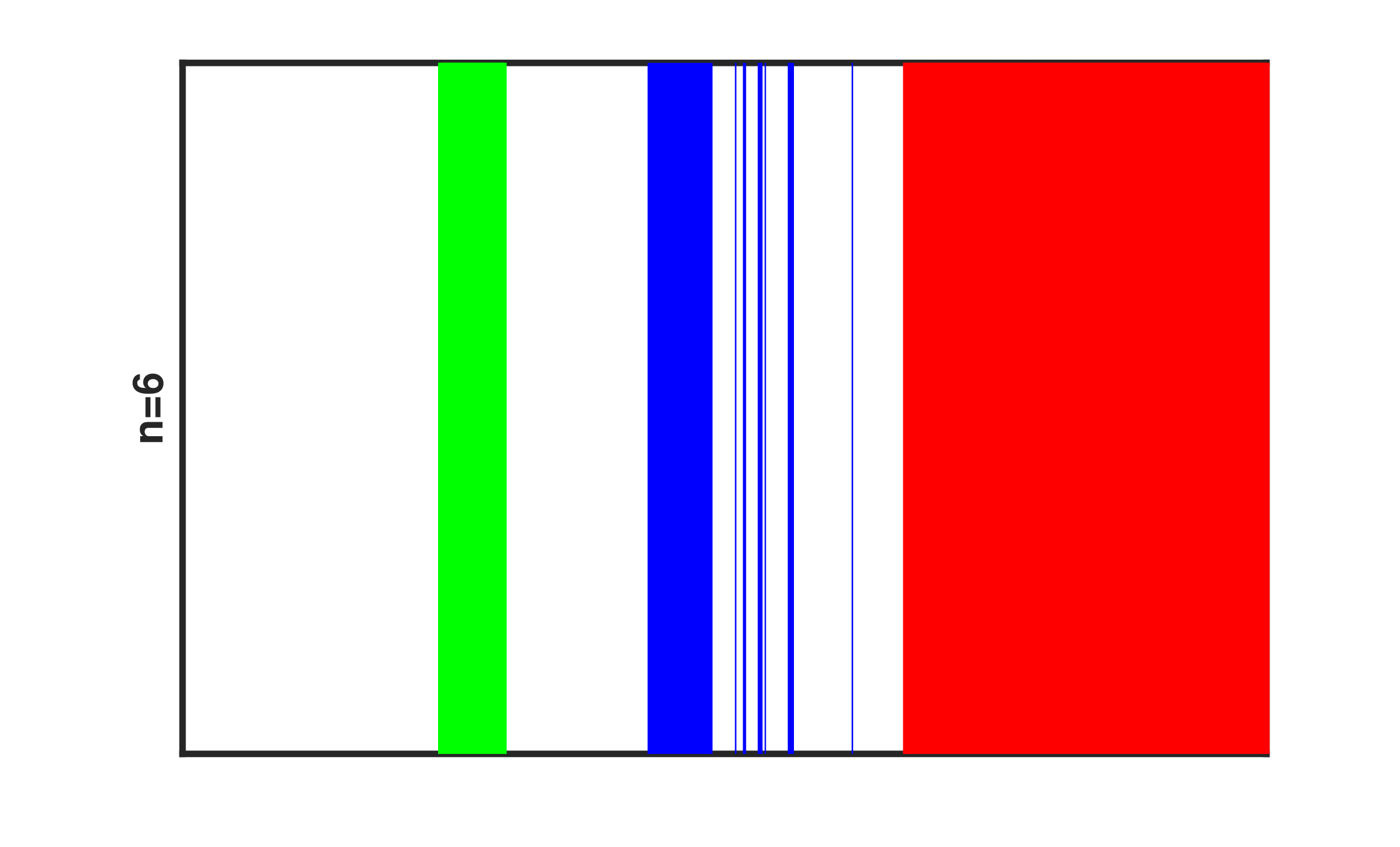}\\
	\includegraphics[{angle=0,width=12cm,height=3cm}]{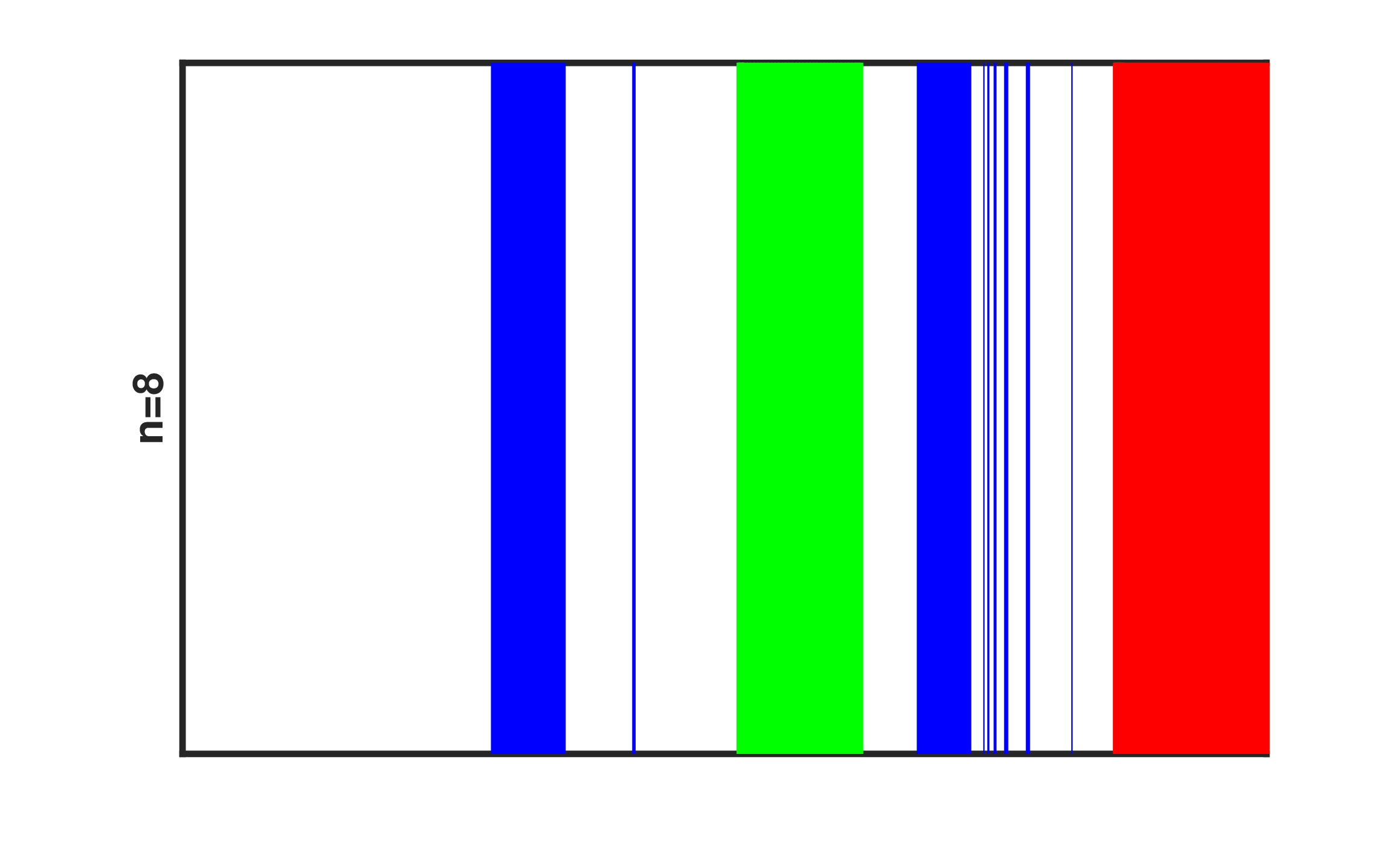}\\
	\includegraphics[{angle=0,width=12cm,height=3cm}]{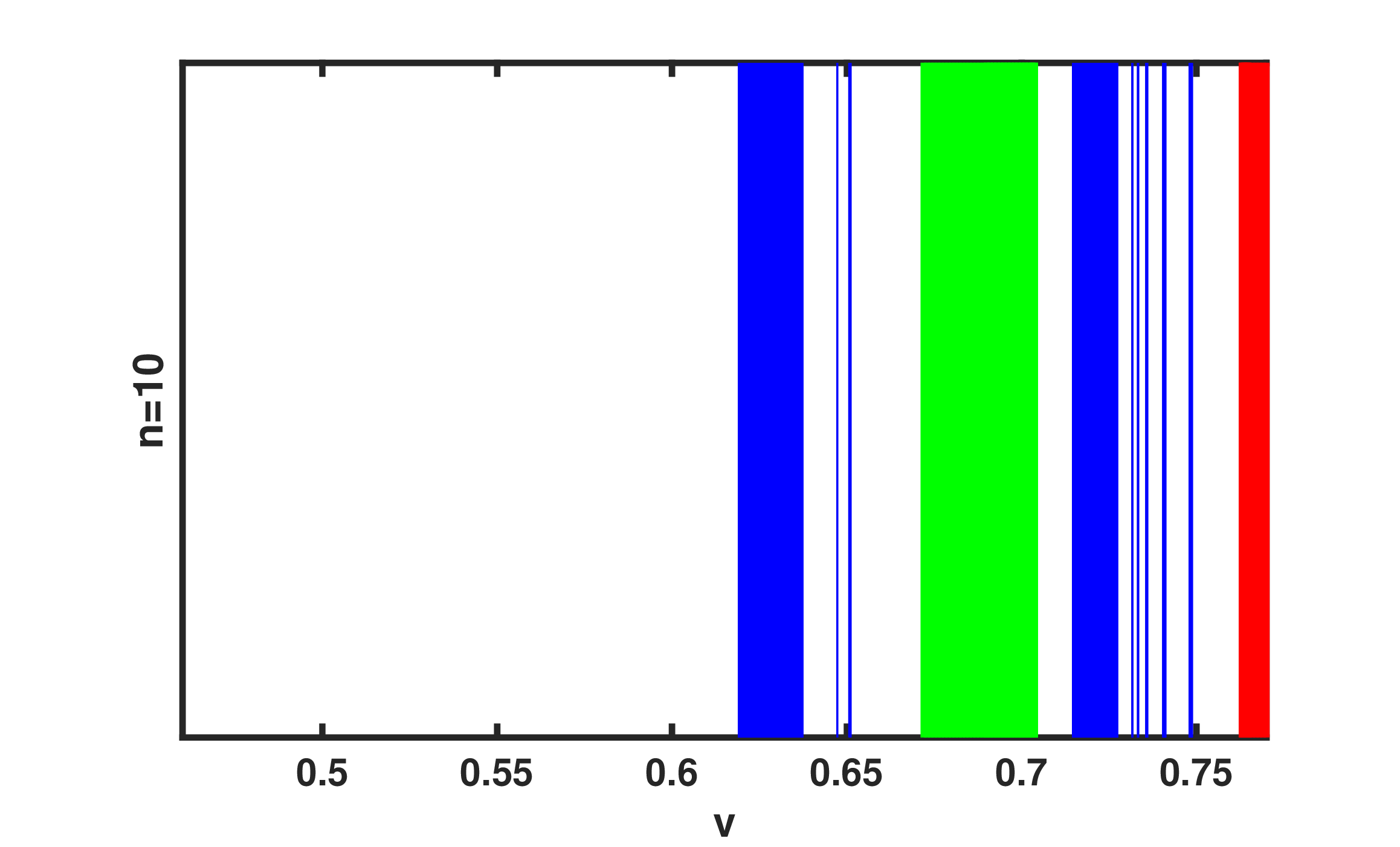}
	\caption{Antikink-kink collisions: bion (white), two propagating oscillations (blue), two-bounce windows (green) and inelastic scattering with one-bounce (red) for (a)$n=4$, (b)  $n=6$, (c)  $n=8$,  and (d) $n=10$.}
	\label{oscwin}
\end{figure}

\begin{figure}
	\includegraphics[{angle=0,width=8cm,height=5cm}]{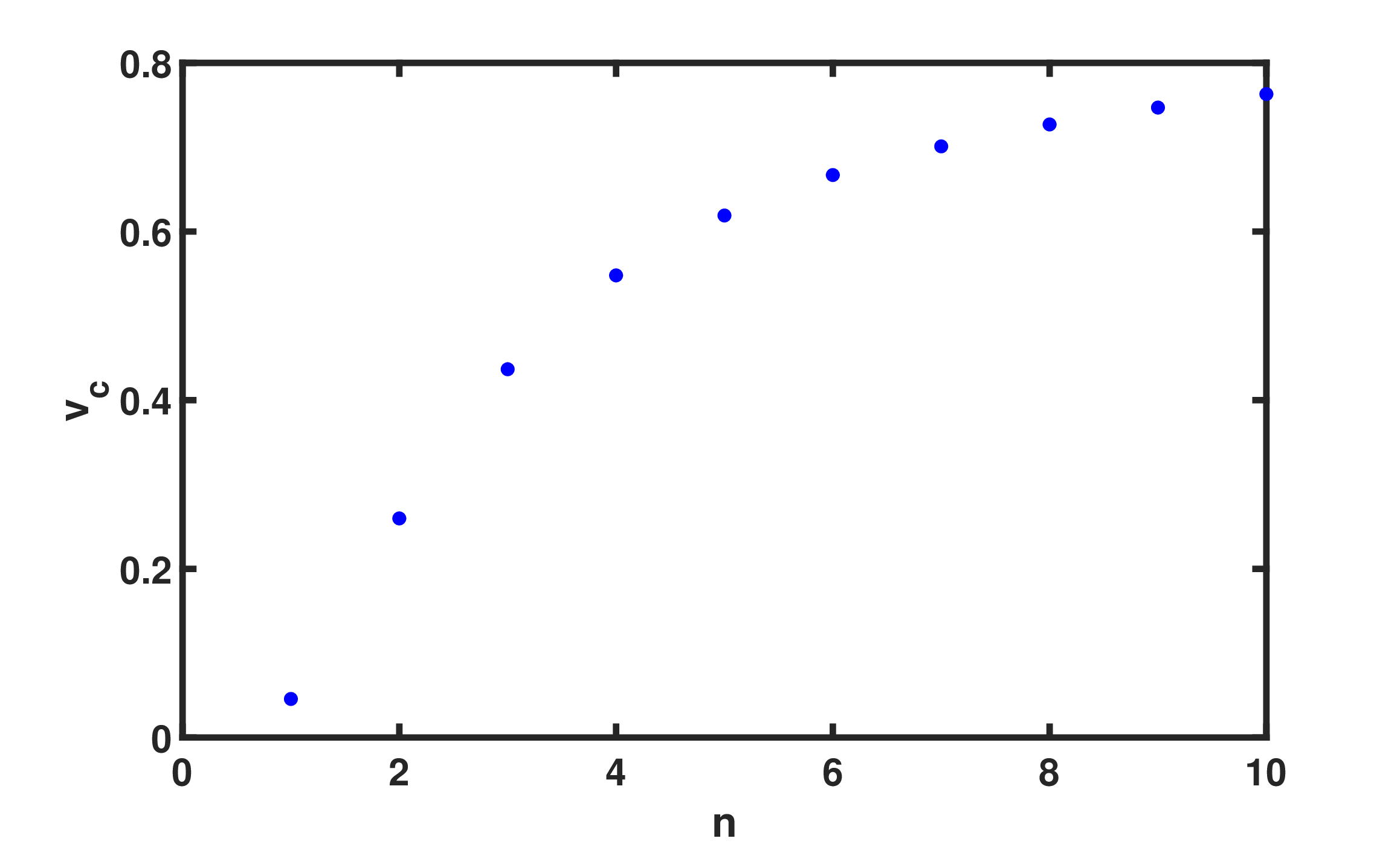}
	\caption{Antikink-kink collision: critical velocity as a function of $n$.}
	\label{velcriAK}
\end{figure}

The antikink-kink scattering can also lead to the production of propagating oscillations. One example of this is  presented in the Fig. \ref{oscillon1} for $n=6$. The structure of the antikink-kink scattering as a function of $v$ and for several values of $n$ is summarized in the Fig. \ref{oscwin}. There one can see the suppression of most of the two-bounce widows (green regions in the Fig. \ref{oscwin}) with the increase of $n$. Also, one can see that for $v>v_c$ there is inelastic scattering of the antikink-kink pair, and that $v_c$ increases with $n$ (shown explicitly in the Fig. \ref{velcriAK}). The process is characterized by the scalar field at the center of mass bouncing once (red region in the Fig. \ref{oscwin}). The region of production of two propagating oscillations is identified in the diagrams in blue. The structure of the windows is intrincate, with no clear pattern for low $n$. 
For $n\geq 8$ one can see some similarity in the structure of the diagrams, with the number of windows being preserved. We note the windows being pushed to larger velocities for larger values of $n$ (compare the Figs. \ref{oscwin}c and \ref{oscwin}d).


\section { Conclusion}


In this work we considered a class of models that depends on a single parameter $n$, which engenders an interesting property: for $n=1$, it is the well-known $\phi^6$ model, and as $n$ changes to higher integers, it supports kinklike configurations that tend to acquire a semi-compact profile when $n$ is very large. Due to the richness of the solutions, the effects of $n$ was investigated both in the kink-antikink and in the antikink-kink collisions.

As we have shown, the kink-antikink scattering is characterized by the production of bion states and propagating oscillations. The propagating oscillations are produced in windows of velocities that increase in number and thickness with the increase of $n$. This effect was also observed, but in lower extension, for antikink-kink scattering. There, beside bion states and propagating oscillations (in an almost constant number for $n\geq 8$), one can have two-bounce windows in a reduced number for large $n$. The pattern of the diagrams that describe the scattering (Figs. \ref{oscka1}, \ref{oscka2} and \ref{oscwin}) is intrincate, but the richness of the diagrams for large $n$ that describe oscillations for kink-antikink scattering (Figs. \ref{oscka1} and \ref{oscka2}) can be related to the presence of more vibrational states for the kink-antikink pair (as seen in the Fig. \ref{vibra}a). Also, the similarity for large $n$ of the structure of the diagrams for antikink-kink scattering (see the Fig. \ref{oscwin}) can be related to the almost constant squared frequencies of the vibrational states (cf. Fig. \ref{fig3}). That is, the linear stability analysis corroborates the observation that a defect with a semi-compacton character present a rich pattern of scattering.

A result of particular interest is that the semi-compactness character of the kinklike configuration for larger values of $n$ is important for the production of multiple oscillations, as it is evident from the richer structure of the kink-antikink scattering in comparison with the antikink-kink collision. Indeed, as shown in the Sect. II, each solution is asymmetric (except for the $n=2$ case), with a kink profile on one side and an (almost) compact profile one on the other side, as we increase $n$ to larger and larger values. For this reason, in the antikink-kink collision the left and right tails of the composed structure have the standard kinklike profile. This does not occur for the kink-antikink collision, since for larger values of $n$, the left and right tails of the composed structure engender the compact profile. As we have shown, the difference appears as very distinct outcomes of the two collisions, and the results may be of current physical interest. As we have shown in Eq. \eqref{exp}, the tail of the compact portion of the solution has a very short double exponential range, which acts to modify the results of the collision. This is somehow similar to the recent studies described in \cite{chri1,chri}, which also found differences between kink and antikink scatterings, although there the authors considered kinklike configurations with long range interactions. In this sense, our results seems to close a gap, showing that asymmetric kinks produce distinct outcomes when they collide: there are different collisions in the case of asymmetric kinks with standard exponential interaction as in the $\phi^6$ model \cite{dorey1}, with the long range polynomial interactions considered in \cite{chri1} for higher power in the scalar field, and also with very short double exponential tails which we investigated in the present work. The collisions described in \cite{dorey1}, \cite{chri1} and in the present work differ from one another, and can be used to describe different situations of current physical interest.


\section{Acknowledgements}

D.B. acknowledges CNPq (Grants No. 303469/2019-6 and No. 404913/2018-0) and Paraiba State Research Foundation (Grant 0015/2019) for financial support.  A.R.G thanks CNPq (brazilian agency) through grants 437923/2018-5 and 311501/2018-4 for financial support. This study was financed in part by the Coordena\c c\~ao de Aperfei\c coamento de Pessoal de N\'ivel Superior - Brasil (CAPES) - Finance Code 001. F.C.S. and A.R.G thank FAPEMA - Funda\c c\~ao de Amparo \`a Pesquisa e ao Desenvolvimento do Maranh\~ao through grants PRONEM 01852/14, Universal 00920/19, 01191/16 and 01441/18.



\end{document}